\documentclass[a4paper,fleqn,usenatbib]{mnras}

\usepackage{newtxtext,newtxmath}

\usepackage[T1]{fontenc}
\usepackage{ae,aecompl}
\usepackage{subfig}
\usepackage{color}

\newcommand{\brunt}{Brunt--V\"ais\"al\"a}

\usepackage{graphicx}	
\usepackage{amsmath}	
\usepackage{amssymb}	

\title[Onset of Nonlinear IGWs]{Onset of Nonlinear Internal Gravity Waves in Intermediate-Mass Stars}

\author[Ratnasingam et al.]{
R.~P. Ratnasingam,$^{1}$\thanks{r.p.ratnasingam2@newcastle.ac.uk}
P.~V.~F. Edelmann,$^{1}$
T.~M. Rogers$^{1,2}$
\\
$^{1}$School of Mathematics, Statistics and Physics, Newcastle University, Newcastle upon Tyne, NE1 7RU, UK\\
$^{2}$Planetary Science Institute, Tucson, AZ 85721, USA\\
}

\date{Accepted XXX. Received YYY; in original form ZZZ}

\pubyear{2018}

\begin{document}
\label{firstpage}
\pagerange{\pageref{firstpage}--\pageref{lastpage}}
\maketitle

\begin{abstract}
Internal gravity waves (IGW) propagate in the radiation zones of all stars. During propagation, their amplitudes are affected by two main features: radiative diffusion and density stratification. We have studied the implications of these two features on waves traveling within the radiative zones of non-rotating stars with stellar parameters obtained from the one dimensional stellar evolution code, MESA. As a simple measure of induced wave dynamics, we define a criterion to see if waves can become nonlinear and if so, under what conditions. This was done to understand the role IGW may play in angular momentum transport and mixing within stellar interiors. We find that the IGW generation spectrum, convective velocities and the strength of density stratification all play major roles in whether waves become nonlinear. With increasing stellar mass, there is an increasing trend in nonlinear wave energies. The trends with different metallicities and ages depend on the generation spectrum.
\end{abstract}


\section{Introduction}
Internal gravity wave propagation in stably stratified environments is a widely studied subject as these waves occur naturally in Earth's atmosphere and oceans \citep{2002AnRFM..34..559S,2003RvGeo..41.1003F,sutherland2010internal}, and are an important factor in the transport of material, angular momentum and energy. For instance, atmospheric gravity waves are formed when a uniform layer of air encounters an obstacle such as a mountain, in a stratified atmospheric region. These waves can then carry angular momentum and energy upwards towards the ionosphere, leading to the formation of wave clouds. Internal gravity waves also impose themselves on ocean surfaces, leading to formation of periodic calm and rough patches which are visible in satellite images. 

In stars, IGW propagate only in radiation zones and as they propagate, some of them are reflected near stellar surfaces, leading to constructive interference and the formation of standing waves. The detection of these g-modes has been one of the central aims of asteroseismology \citep{aerts2010asteroseismology}. All propagating gravity waves undergo radiative damping \citep{1993A&A...279..431S,1997A&A...322..320Z}. Some of them eventually break \citep{2010MNRAS.404.1849B} or encounter critical layers. Damping occurs due to radiative diffusion of photons across the IGW wavelength where perturbation temperatures have opposite signs. Wave breaking, on the other hand, is a non-linear process that occurs when the wave amplitude reaches a critical value, which causes the energy in waves to be converted to turbulent kinetic energy. These processes lead to energy, angular momentum transport and chemical mixing within the radiation zone \citep{1981ApJ...245..286P,1991ApJ...377..268G,1994A&A...281..421M,2005Sci...309.2189C,2017arXiv170904920R}. In most cases, IGWs are excited stochastically by turbulent convection within the stellar convection zone, leading to a predominantly inward (towards the center) propagation \citep{2002ApJ...570..825B,2003Ap&SS.284..237D} for stars with convective envelopes ($M < 1.5\,M_\odot$) and predominantly outward propagation (towards the surface) for stars with convective cores ($M > 1.5\,M_\odot$).   

Internal gravity waves have been proposed to solve a host of unexplained observations in stars. \cite{1981ApJ...245..286P} first invoked IGW to explain what was then the neutrino problem. \cite{1991ApJ...377..268G} and a series of papers by Montalban and Schatzman \citep{1994A&A...281..421M,1996A&A...305..513M,2000A&A...354..943M} suggested IGW could produce the enhanced mixing \citep{1981ApJ...248..751P} needed to explain the observed lithium depletion in F stars. \cite{1997A&A...322..320Z} and later, \cite{1999ApJ...520..859K} suggested that IGW could explain the uniform rotation of the solar radiative interior. \cite{2003MNRAS.340..722D} showed that turbulent mixing induced by IGWs can mix protons into helium and carbon rich zones causing the formation of carbon-12 pockets in low-mass asymptotic giant branch (AGB) stars. \cite{2012ApJ...758L...6R} suggested IGW could change the surface rotation rate of intermediate-mass stars, thus explaining the discrepancy of obliquity between hot Jupiters around hot and cool stars. \cite{2009A&A...508..409A} and \cite{2015ApJ...806L..33A} suggested IGW could explain the observed macroturbulence seen across the upper main sequence. \cite{2014ApJ...796...17F} suggested IGW could cause the enhanced angular momentum transport needed to explain the observed differential rotation in evolved stars and may be responsible for pre-supernovae mass ejection \citep{2015ApJ...810..101F}.  Although IGW may be responsible for a variety of effects in stellar and planetary interiors, the details of their influence depends sensitively on the spectrum of waves generated by turbulent convection, which is still uncertain.

Neglecting the details of wave generation and assuming that wave propagation is linear and no critical layers form, wave amplitudes depend only on density stratification and thermal diffusivity. While very few theories have considered the possibility of wave breaking, it may be possible in some circumstances \citep{2010MNRAS.404.1849B,2013ApJ...772...21R}. In stars with convective cores, IGWs propagate outward, into a region of decreasing density, causing the wave amplitudes to increase (in the absence of thermal diffusion), which ultimately means these waves are likely to be more dynamically relevant in intermediate-mass stars than lower-mass stars. Hence, we focus our study here on intermediate-mass stars. While numerical simulations, such as those carried out in \cite{2013ApJ...772...21R} can simulate the convective-radiative interface and the self-consistent generation of waves, they do so at artificially high viscosities, high thermal diffusivities and/or reduced dimensionality.  Therefore, it is currently impossible to self-consistently simulate the generation and propagation of IGWs within a realistic stellar interior.  

In this paper, we take a different approach. Assuming a given generation spectrum and that the waves propagate linearly through the radiation zones of stars, we determine what the likelihood is that these waves could become nonlinear. We use realistic thermal diffusivities and other stellar properties generated by the one-dimensional stellar evolution code, Modules for Experiments in Stellar Astrophysics (MESA, version 8845; \cite{2011ApJS..192....3P}). To investigate how the wave amplitude changes with radius in a star, we consider two main effects: amplitude enhancement due to density stratification and amplitude reduction due to radiative diffusion. We then calculate the nonlinearity parameter as a function of wavenumber, frequency and initial amplitude to determine the likelihood of wave breaking. The work done in \cite{2013MNRAS.430.1736S}, which predicted the observability of convectively-driven g-modes under the influence of radiative damping in stars with convective cores, is similar, but here, we focus on the possibility of gravity waves becoming nonlinear. While this approach investigates the amplitude evolution of IGWs, it does not look into the influence of these breaking waves on angular momentum transport and chemical mixing.

This paper is organised as follows. In section \ref{sec:IGW_pheno}, we discuss IGW physics such as generation by turbulent convection, wave propagation in the linear regime and IGW nonlinearity. In section \ref{sec:Model_setup}, we discuss the stellar models generated by MESA and the initial parameters used to set up these models. In section \ref{sec:results}, we present the analysis done on wave propagation and nonlinearity for varying masses and ages. Section \ref{sec:dis} discusses these results and concludes this work.

\section{Internal Gravity Wave Phenomenon}\label{sec:IGW_pheno}
\subsection{IGW Generation by Turbulent Convection\label{sec:IGW_gen}}

The classic prescription for wave generation by convection goes back to \cite{1990ApJ...363..694G}, which was originally developed for pressure waves (p-modes). This prescription was then extended to gravity waves by \cite{1999ApJ...520..859K}. The original prescription treats wave generation by bulk convection due to three sources: changes in entropy at fixed pressure (monopole terms), buoyancy variations (dipole terms) and internal stresses (quadrupole terms). In that work, it was argued that while the monopole and dipole terms generate more acoustic radiation, they very nearly cancel each other and therefore, the only term considered was the quadrupole term, or the term due to internal Reynolds stresses. This assumption was carried forward in \cite{1999ApJ...520..859K}, yet it is unclear that this approximation is valid for IGW. Although there is evidence that treating IGW generation as an inhomogeneous wave equation with a source term is appropriate \citep{2013MNRAS.430.2363L}, it is still unclear what this source term should be or whether the theoretical picture is an accurate one. 

The most obvious deficiency of this model is that it does not account for wave generation by overshooting plumes \citep{rieutord1995a} while it is well known from experiments \citep{ansong_sutherland_2010} and simulations that plume generation is relevant and possibly dominant. In terms of theoretical work, this was first studied by \citet{townsend1966a} in a terrestrial context. \citet{montalban2000a} used a plume model developed by \citet{rieutord1995a} to derive the IGW spectrum generated at the bottom of the convection zone of a solar-type star. Later, \citet{pincon2016a} developed a semi analytical model for the same situation to be used in 1D stellar evolution codes. A common result of all these models is an exponential frequency spectrum, proportional to $\exp\left(-\omega^2 t_\mathrm{b}^2\right)$ with a characteristic plume incursion timescale~$t_\mathrm{b}$ and $\omega$ being the generated wave frequency. The wave number dependence is similarly exponential. Then, the total spectrum results from a combination of plumes of different sizes and timescales. Although we believe this generation spectra to be important, we do not include it here but note that the numerical generation spectra from \cite{2013ApJ...772...21R} is likely to be similar. 

To date, no numerical simulations have been able to reproduce any of the theoretically predicted wave spectra \citep{2013ApJ...772...21R,2014A&A...565A..42A,2015A&A...581A.112A} and similarly, neither have any laboratory experiments \citep{ansong_sutherland_2010}. However, a recent work by \cite{CoustonLouis-Alexandre2018Tefs} has confirmed one of theoretical generation spectra investigated in this work through numerical simulations involving the Boussinesq approximation. Also, there is new evidence from 3D simulations that numerical solutions are similar to plume spectra for a variety of plume length and timescales (Edelmann at al (submitted)). Whether this discrepancy between simulations/experiments and theory is due to inadequate assumptions of the theory or unrealistic parameters of the simulations, or both, is still not known. Therefore, the spectrum of waves generated by convection remains a fundamental problem in all theories of wave transport (of angular momentum, energy or species).  
The first wave generation spectrum explored in this work is that from \cite{1999ApJ...520..859K}. Using flux equations calculated for p-modes from previous work \citep{1990ApJ...363..694G,1994ApJ...424..466G}, \cite{1999ApJ...520..859K} assumed continuity of radial velocities across the convective-radiative interface and derived a flux equation for IGW generated by Reynolds stresses as
\begin{equation}\label{eq:kumar_flux}
F \sim  u_c^3 \; \frac{\rho_c k_h^3}{r_c^2 N_c} \left(\frac{u_c}{\omega_c}\right)^5\left(\frac{\omega}{\omega_c}\right)^{-10/3}\exp\left[-k_h^2 \left( \frac{u_c}{\omega_c}\right)^2 \left( \frac{\omega}{\omega_c} \right)^{-4/3}\right],
\end{equation}
where $\ell$ is the spherical harmonic degree (we will be referring to $\ell$ as wave number, unless stated otherwise) while $u_c$, $\omega_c$, and $\rho_c$ are the bulk convective velocity, convective turnover frequency (in units of Hz) and density at the convective-radiative boundary, respectively. The term, $N_c$ is the \brunt{} frequency at the bottom of the radiation zone while $r_c$ is the thickness of the convection zone. The horizontal wavenumber, $k_h$ is defined as
\begin{equation}\label{eqn:kh}
k_h = \frac{\sqrt{\ell~(\ell +1)}}{r}.
\end{equation}
From here on, the subscript $h$ will represent horizontal quantities while $v$ will represent vertical quantities. We convert Eq.~\eqref{eq:kumar_flux} into a displacement equation by equating it to $\rho_c \left(\omega \xi_{h}\sqrt{\ell(\ell + 1)} \right)^2 v_g$ as given in \cite{1999ApJ...520..859K}, where $v_g$ is the vertical IGW group velocity (see Eq.~\eqref{groupv}), $\xi_h$ is the horizontal wave displacement and we have assumed that $\xi_h \gg \xi_v$. Using $u_h = \xi_h\omega$ gives the following relation,
\begin{equation}\label{eqn:kumar}
\begin{aligned}
u_{h,K} \sim u_c \frac{ k_h^2}{r_c \sqrt{\ell(\ell + 1)}}  \left( \frac{u_c}{\omega_c}\right)^{3} \left( \frac{\omega}{\omega_c}\right)^{-13/6} \\ \exp \left[- \frac{k_h^2}{2} \left( \frac{u_c}{\omega_c}\right)^2\left( \frac{\omega}{\omega_c}\right)^{-4/3} \right], 
\end{aligned}
\end{equation}
where $u_h$ is the horizontal fluid velocity and the subscript K refers to ``Kumar". The work by \cite{1999ApJ...520..859K} was revisited by \cite{2013MNRAS.430.2363L}, who also produced theoretical predictions for the IGW generation flux, but for three different stratification profiles at the convective-radiative transition. All three profiles have various power laws on $k_{h}$ and $\omega$, which are determined by the \brunt{} frequency profile across the convective-radiative interface (i.e.\ discontinuous, smooth continuous or piecewise linear). Assuming $N \gg \omega$, the horizontal velocity perturbation can be written as
\begin{equation} \label{eq:LD_final}
u_{h,LD} \sim u_c \left( \frac{k_h u_c}{\omega_c}\right)^{5/2}\left( \frac{\omega}{\omega_c}\right)^{-17/4},
\end{equation}
where the subscript LD refers to ``Lecoanet Discontinuous". We only work with the discontinuous case here, as this scenario is currently considered the most probable in stars (Lecoanet, private communication). 

In addition to theoretical work, there has been significant numerical work done to investigate the transition between convective and radiative regions and convective overshoot \citep{1986ApJ...311..563H,2002ApJ...570..825B} with a few having focused on IGW generation \citep{2005MNRAS.364.1135R,2011ApJ...742...79B,2013ApJ...772...21R,2014A&A...565A..42A,2015A&A...581A.112A}. For instance, \cite{2013ApJ...772...21R} predicts that for a star with zero rotation, the energy carried\footnote{Approximately 79\% of the energy in IGWs were fit by a separable function, where E($\omega,l$) $\propto$ g($\omega$)*f($\ell$).} by IGW scales as $E \propto \omega^{-1.2} \left(\sqrt{\ell(\ell + 1)}\right)^{-1.8}$ in the low wave number ($\ell \lesssim 10$) and low frequency ($\omega \lesssim 10\;\mu$Hz) regime, just outside the convection zone. While it is possible that the negative exponent associated with the wavenumber dependence is due to the reduced dimensionality of those simulations, more recent three-dimensional simulations also show reduced efficiency for shorter wavelengths/larger wavenumbers (Edelmann at al (submitted), \cite{2014A&A...565A..42A}). Furthemore, these simulations show a similar frequency dependence, so we take the 2D simulations to be representative. Using $E \propto \left(\omega \xi_{h}\right)^2\propto u_{h,R}^2$ where $\xi$ is the fluid displacement amplitude and the subscript R refers to ``Rogers", $u_{h,R}$ is found to be proportional to $\omega^{-0.6} \left(\sqrt{\ell(\ell + 1)}\right)^{-0.9}$.

The IGW generation spectrum, which represents the collection of waves generated at the convective-radiative interface can be summarised as power laws in both wave number and frequency and described by:
\begin{equation}\label{eq:spectra_proportionality}
u_0(\omega,\ell) \propto \omega^m \left(\sqrt{\ell(\ell + 1)}\right)^n
\end{equation}
where the values of $m$ and $n$ for different prescriptions are shown in Table \ref{table:mandn}. An additional IGW generation spectrum, the flat spectrum, has been introduced in Table \ref{table:mandn}, which represents a spectrum with no dependence on $\ell$ and $\omega$. This spectrum will act as a test case and will be hereon referred to as spectrum F. Also, the table does not include the extra dependences of spectrum K on $k_h$ and $\omega$ in the exponential term. However, we do consider the effect of this exponential term in our analysis.
 
\begin{table}
	\centering		
	\begin{tabular}{ccc}
		\hline \hline
		Spectra & m & n \\ \hline \hline 
		Flat[F] & 0 & 0 \\
		\cite{1999ApJ...520..859K}[K] & -2.17 & 1 \\
		\cite{2013MNRAS.430.2363L}[LD] & -4.25 & 2.5 \\
		\cite{2013ApJ...772...21R}[R] & -0.6 & -0.9 \\ \hline
	\end{tabular}
	\caption{The table shows how $m$ and $n$ in Eq.~\eqref{eq:spectra_proportionality},  are defined for different works. The flat spectrum represents one with no dependence on $\omega$ or $\ell$. The letters shown in square brackets will be used to represent its respective spectrum.}
	\label{table:mandn}
\end{table}

\subsection{Radiative Damping and Density Stratification}
As IGW generated at the convective-radiative interface propagate outward, their amplitudes are affected by two main features; radiative damping and density stratification. We follow the work of \cite{1999ApJ...520..859K}, which built on the work of \cite{1981ApJ...245..286P}, to investigate the effect of radiative damping on wave amplitudes. The damping opacity, $\tau$ is defined as 
\begin{equation}\label{tau}
\tau (\omega,\ell,r) = \int_{r_\text{interface}}^{r} \frac{\gamma [\omega,\ell,r']}{\left|v_g [\omega,\ell,r']\right|} dr'
\end{equation}
where $\gamma$ is the damping rate, given by
\begin{equation}\label{gamma}
\gamma(\omega,\ell,r) = Kk_v^2
\end{equation}
and $k_v$, the vertical wavenumber, is defined as
\begin{equation}\label{eq:kv_to_kh}
k_v^2 = k_h^2\frac{N^2-\omega^2}{\omega^2},
\end{equation}
where $N$ is the \brunt{} frequency. In MESA, $N$ is defined using the Ledoux criterion as
\begin{equation}\label{eq:bvf}
N^2 = \frac{g^2 \rho}{P}\frac{\chi_T}{\chi_{\rho}}(\nabla_{\mathrm{ad}} - \nabla_{T} + B),
\end{equation}
where $g$ is the gravitational acceleration, $\chi_T$ is $(\partial \ln P/\partial \ln T)_{\rho}$, $\chi_{\rho}$ is $(\partial \ln P/\partial \ln \rho)_{T}$, $\nabla_{\mathrm{ad}}$ is the adiabatic temperature gradient, $\nabla_{T}$ is the true temperature gradient and $B$ is the compositional gradient. In the case of ideal gas, ${\chi_T}/{\chi_{\rho}} = 1$. Note that it is common to work in the asymptotic regime and assume $ N \gg \omega$. However, we keep the $\sqrt{N^2/\omega^2 -1}$ term in our propagation equation, mainly because although the asymptotic assumption applies in most of the radiation zone, it fails to do so close to the convective-radiative boundaries as $N$ approaches 0.
The vertical group velocity, $v_g$ is defined as
\begin{equation}\label{groupv}
v_g(r) = \frac{\partial\omega}{\partial k_v}=-\frac{(N^2-\omega^2)^{1/2}\omega^2}{k_hN^2}.
\end{equation}
The thermal diffusivity, $K$, is given by 
\begin{equation} \label{eqn:Tdiff}
K(r) = \frac{16\sigma {T}^3}{3{\rho}^2 {\kappa} {c}_p},
\end{equation}
where $\sigma$ is the Stefan-Boltzmann constant while $T$, $\rho$, $\kappa$ and $c_p$ are stellar temperature, density, opacity and specific heat capacity, respectively, which are all functions of stellar radius. The full expression of terms inside the integral of Eq.~\eqref{tau} then, becomes
\begin{equation} \label{eq:gam_over_vg}
\frac{\gamma}{v_g} = \frac{16\sigma T^3}{3\rho^2 \kappa c_p}\left(\frac{\left(\ell(\ell + 1)\right)^{3/2} N^3}{r^3 \omega^4}\right)\left(1-\frac{\omega^2}{N^2}\right)^{1/2}.
\end{equation}
Using stellar parameters and \brunt{} frequencies obtained directly from MESA, $\gamma (r)/ v_g (r)$ was calculated and integrated using the cumulative trapezoidal method. Values of $\tau$ (see Eq.~\eqref{tau}) for different propagation distances, ($r - r_\text{interface}$), were then inserted into the following equation to obtain wave amplitude as a function of radius:
\begin{equation} \label{eq:propeqn}
u_h(\omega,\ell,r) = u_0(\omega,\ell) \left( \frac{r_0}{r}\right)^{3/2}\left(\frac{\rho_c}{\rho(r)}\right)^{1/2} \left(\frac{N^2 - \omega^2}{N_c^2 - \omega^2}\right)^{1/4} e^{-\tau/2},
\end{equation}
where $u_0$ is the velocity perturbation amplitude (which takes the magnitude of convective velocities, $u_c$, into account) from different generation spectra described in Section~\ref{sec:IGW_gen} and $r_0$ is the radius at which the waves are generated. This expression, excluding the damping term, is within the WKB approximation. It can be derived from the anelastic, linearised continuity, momentum and energy equations, as done in \cite{1981ApJ...245..286P} and \cite{1997A&A...322..320Z}, with the assumptions of adiabatic perturbations and zero source terms in a medium with density stratification.

\subsection{Nonlinearity Parameter}
\cite{phillips1966dynamics}, \cite{1981ApJ...245..286P} and \cite{2010MNRAS.404.1849B} have shown that the nonlinearity of waves can be represented by the ratio of wave displacement to wavelength in either horizontal or vertical directions. This leads to the definition of the nonlinearity parameter, $\epsilon$, as follows:  
\begin{equation} \label{eqn:WBcond2}
\epsilon=\xi_{h} k_{h} = \frac{u_{h}}{\omega} k_{h}, 
\end{equation}
where we have replaced the horizontal displacement with horizontal perturbation velocity over frequency. An IGW is said to be nonlinear, when its displacement becomes comparable to or larger than its wavelength, which is equivalent to the strong condition, $\epsilon \geqslant 1$. However, one also expects that as $\epsilon$ approaches 1, nonlinear effects can become important.

To consider only the horizontal components, we can first show that 
\begin{equation}\label{eqn:amp_h}
\frac{k_h}{k_v}=\sqrt{\frac{\omega^2}{N^2-\omega^2}},
\end{equation}
from rearranging Eq.~\eqref{eq:kv_to_kh}. In the Boussinesq approximation which applies locally, the divergence of the velocity perturbations is zero and thus, $k_h u_h = k_v u_v$. So, the following relation is expected to be satisfied in the radiation zone:
\begin{equation}
k_h \xi_h \sim \left(\sqrt{\frac{\omega^2}{N^2-\omega^2}}\right)k_v \left(\sqrt{\frac{N^2-\omega^2}{\omega^2}}\right) \xi_v \sim k_v \xi_v, 
\end{equation}
which shows that considering wave nonlinearity in the horizontal and vertical directions leads to the same result.  

\begin{figure*}
	\centering
	\includegraphics[trim={0.0cm 0.3cm 0.0cm 0.0cm},clip,width=0.85\textwidth]{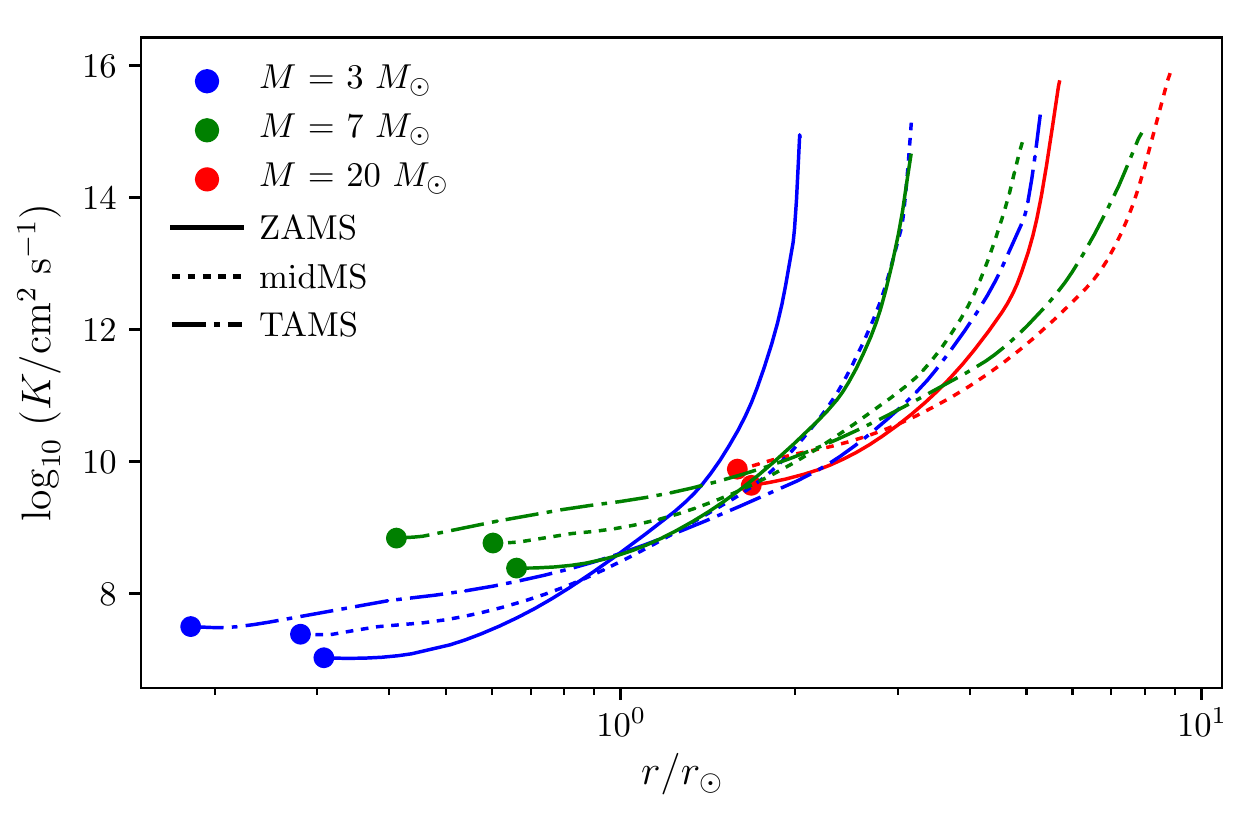}
	\caption{Thermal diffusivity as a function of radius, in units of solar radius for different stellar masses generated using MESA. The value M in the legend represents the mass of the stellar models used in solar mass units. The circular markers at the beginning of each plot indicate the locations of the convective-radiative interface. The large variation in magnitude seen in the thermal diffusivity profiles is mainly due to the large variation in density profiles of stars. \label{fig:Kvsr}}
	
\end{figure*}
\begin{figure}
	\centering
	\includegraphics[trim={0.0cm 0.3cm 0 0.0cm},clip,width=0.9\columnwidth]{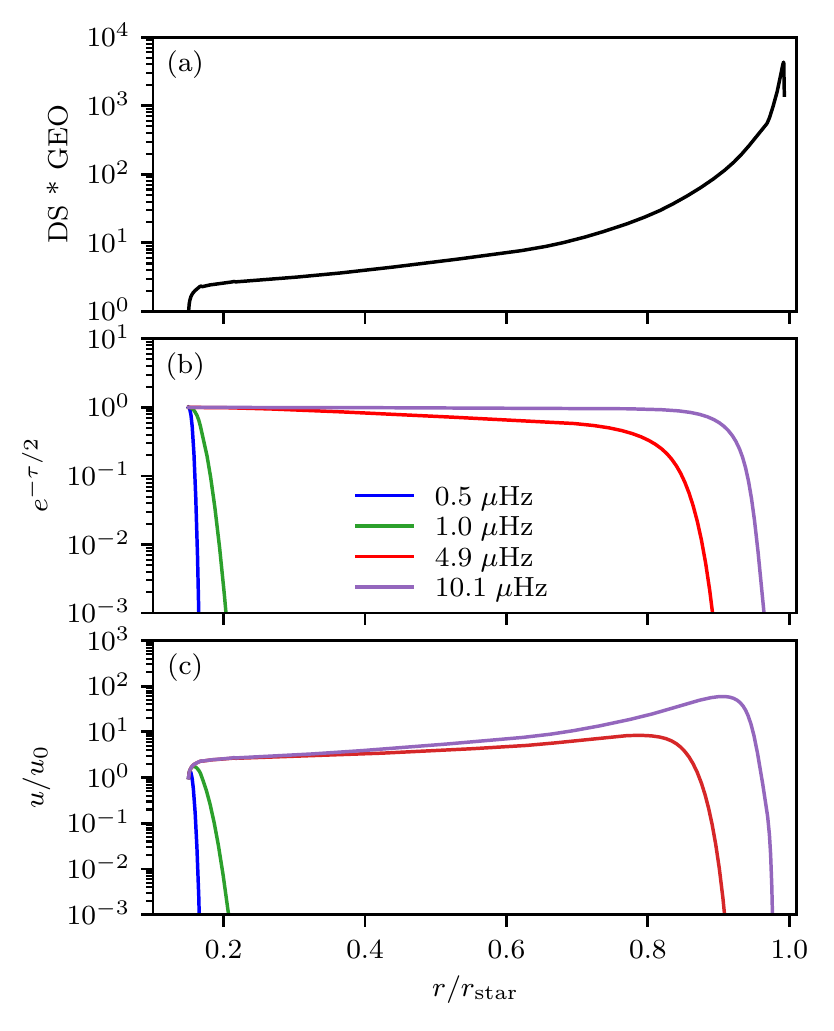}
	\caption{(a) The density stratification term (DS) represented by $\left(\rho_c/\rho\right)^{1/2}\left[(N^2-\omega^2)/(N_c^2-\omega^2)\right]^{1/4}$ for $\omega =$ 10 $\mu$Hz, and the geometric term (GEO) represented by $\left(r_0/r\right)^{1.5}$, (b) radiative damping and the (c) normalised amplitude profiles as functions of radii, in units of total stellar radius. The different coloured lines in the middle and bottom plots represent different IGW frequencies as given in the legend. The wave number, $\ell$ is set to 1.\label{fig:IGW_propagation}}
	\centering
\end{figure}
\begin{figure*}
	\centering
	\includegraphics[trim={0.0cm 0.0cm 0 0.0cm},clip,width=0.9\textwidth]{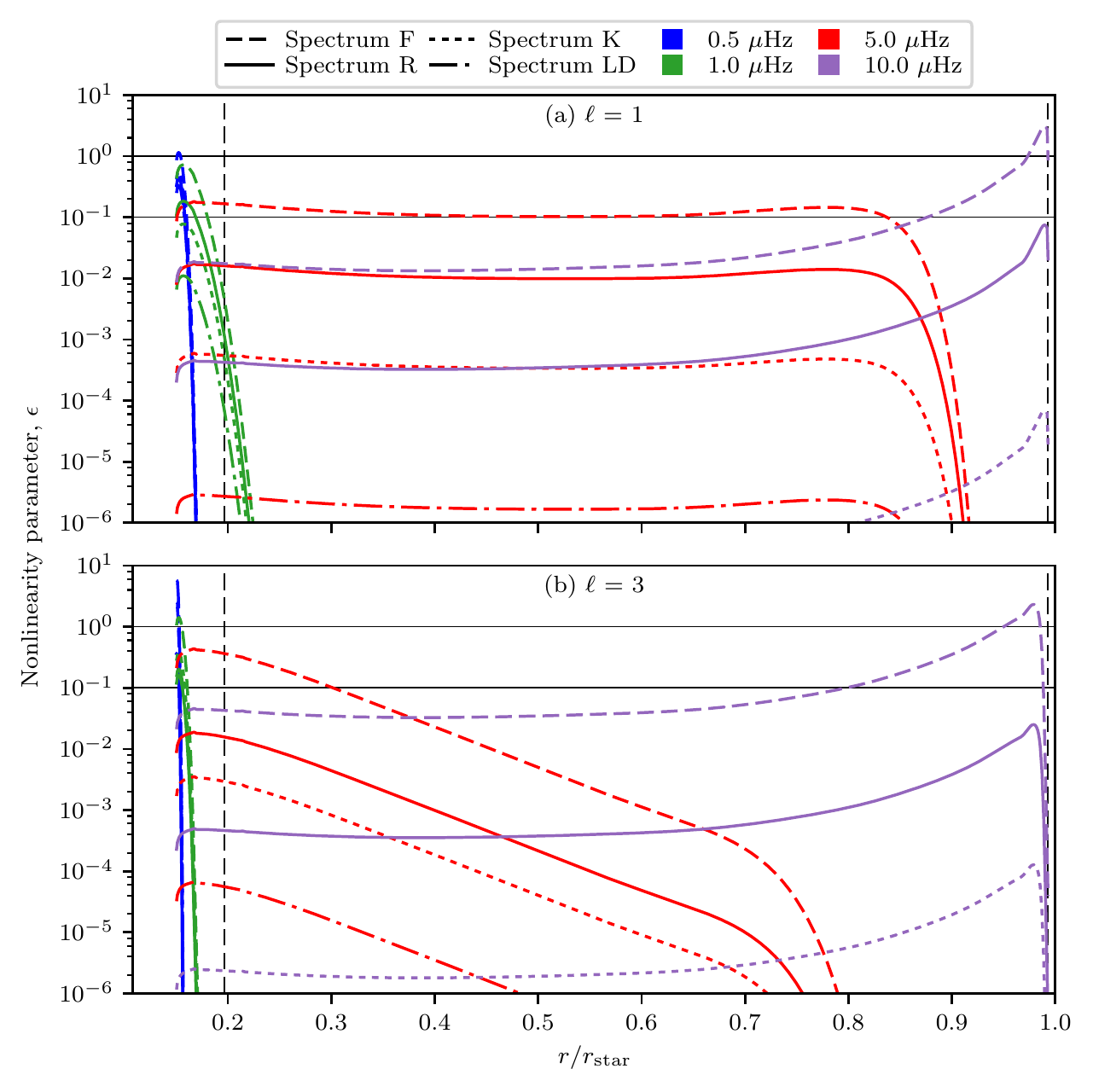}
	\vspace{-12pt}
	\caption{Nonlinearity parameter, $\epsilon$, versus radii for a 3 $M_{\odot}$ star at ZAMS for (a) $l = 1$ and (b) $l = 3$. The two dotted vertical lines show the location of $r_{\mathrm{min,break}}$ and the surface convection zone. while the two straight horizontal lines mark $\epsilon = 1.0$ (top) and $\epsilon = 0.1$ (bottom). The colours blue, green, red and indigo have been used to represent IGW frequencies of 0.5 $\mu$Hz, 1.0 $\mu$Hz, 5.0 $\mu$Hz and 10.0 $\mu$Hz respectively while different linestyles represent different initial spectrum as shown in the legend.  \label{fig:IGW_damping}}
\end{figure*}

\section{Model Setup}\label{sec:Model_setup}
Models of stars with masses ranging from 3 $M_{\odot}$ to 20 $M_{\odot}$ were constructed using the MESA stellar evolution code from ZAMS, when the core hydrogen mass fraction ($X_c$) is approximately 0.7 to middle of main sequence (midMS), when $X_c \approx$  0.35, followed by terminal age main sequence (TAMS), when $X_c \approx$ 0.01. Mass loss was implemented through a fixed stellar wind scheme. Stellar metallicity, $Z$, was set to be equal to the solar value of $Z$ = 0.02, although we do test the effect of varying metallicities in Section~\ref{sec:results_metal}. The mixing length parameter was set 1.8. The sign of the \brunt{} frequency was used to determine the extent of the radiation zone which means 
\begin{equation}
N(r)^2 > 0 
\end{equation}
refers to a radiation zone and
\begin{equation}
N(r)^2 < 0 
\end{equation}
refers to a convection zone. However, to account for the fact that the IGW frequencies must be less than the \brunt{} frequency, the extent of the radiation zone was further limited to regions where the \brunt{} frequency exceeds 100 $\mu$Hz, which was the highest IGW wave frequency we tested. It was found that in all our stellar models, the \brunt{} frequency increases rapidly from zero at the core convective-radiative interface and decreases rapidly back to zero at the surface radiative-convective interface. Therefore, at both the top and bottom convective-radiative boundaries, the regions of the radiative zone cut out by imposing the above condition were small, being less than 0.5\% of the star's total radius (see Table~\ref{table:cutout}).  Thermal diffusivities (see Eq.~\eqref{eqn:Tdiff}) of 8 stellar models were calculated using opacities from MESA as shown in Fig.~\ref{fig:Kvsr}. The 20 $M_{\odot}$ TAMS model has not been included as it was found to develop an intermediate convection zone. It can be seen that for any particular stellar mass, the convection zone shrinks as the star evolves while the radiation zone expands. This is an important factor in IGW propagation which will be addressed further in the Section~\ref{sec:AgeMass}.

\begin{table}
	\centering		
	\begin{tabular}{ccccc}
		\hline \hline
		Stellar Mass & Stellar Age & Inner Cut-out & Outer Cut-out  \\ \hline \hline 
		3 & ZAMS & 0.048 & 0.0 \\
         & midMS  & 0.020 & 0.001 \\ 
         & TAMS  & 0.0 & 0.016 \\ \hline
        7 &  ZAMS & 0.103 & 0.007  \\
		  & midMS  & 0.059 & 0.031 \\ 
          & TAMS  & 0.011 & 0.311 \\ \hline
        20 & ZAMS  & 0.394 & 0.004  \\
		   & midMS  & 0.117 & 0.035  \\ \hline
	\end{tabular}
	\caption{The table shows percentages of the radiative zone (in terms of the total stellar radius) that were cut out close to the inner convective zone (Inner Cut-out) and close to the surface convection zone (Outer Cut-out) for the different stellar models we used.}
	\label{table:cutout}
\end{table}

Figure \ref{fig:IGW_propagation} is an example of how wave amplitude is affected by density stratification and radiative diffusion for a 3 $M_{\odot}$ star at $X_c$ = 0.7. The top panel shows how wave amplitude changes due to density stratification and geometric effects within the radiation zone of the chosen stellar model. Across the radiation zone, density decreases by almost eight orders of magnitude while radius increases by approximately one order of magnitude. This leads to the growth of wave amplitudes as the density stratification term dominates over the geometric term. The trend also shows a sharp increase close to the convective core and a sharp decrease close to the surface convection zone caused by the \brunt{} frequency changing rapidly (by more than three orders of the magnitude). The middle panel shows the effect of radiative diffusion on IGW with an initial wave amplitude of 1 and wave number of 1. It can be observed that waves of higher frequencies are damped less than waves of lower frequencies due to the damping opacity being proportional to $\omega^{-4}$ (see Eq.~\eqref{eq:gam_over_vg}). In stellar interiors, IGWs are affected by both radiative damping and density stratification simultaneously and this can be seen in the bottom panel of Fig. \ref{fig:IGW_propagation}. Waves that do not succumb to damping rapidly (i.e.\ higher frequency waves) undergo an increase in amplitude due to density stratification as they propagate through the radiation zone of a star towards the surface. Close to the surface convection zone, both increasing thermal diffusivity and rapidly decreasing \brunt{} frequency cause IGW amplitudes to decrease rapidly as seen for the case when $\omega$ = 5.0 $\mu$Hz (red line). As the \brunt{} frequency decreases towards the surface, more waves are likely to be internally reflected at radii where their frequencies are equal to the background \brunt{} frequency and possibly set up standing modes. Our analysis does not consider this scenario. Looking at the effects of density stratification and radiative damping alone, one may ask if the wave amplitudes increase enough to become nonlinear before reaching the surface. 

\section{Results}\label{sec:results}
\begin{figure*}
	\centering
	\includegraphics[trim={0.0cm 0.2cm 0 0.0cm},clip,width=0.75\textwidth]{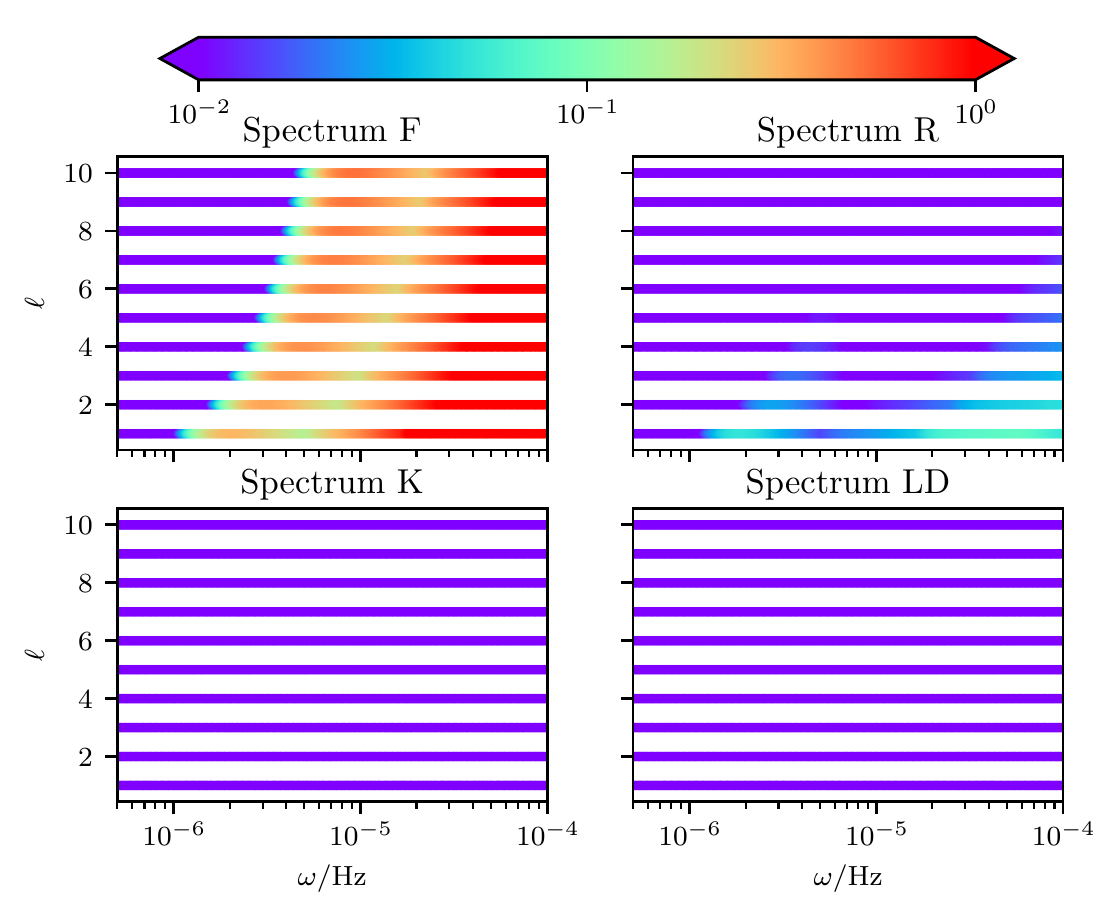}
	\caption{Colour maps of the nonlinearity parameter, $\epsilon$, for different spectra (as described in Table~\ref{table:mandn}). The y-axes represent wave number, $\ell$ and the x-axes represent $\omega$ in Hz. The colours represent the maximum $\epsilon$ reached by an IGW of a given wave number and frequency at radii more 0.5 $H_P$ from the convective-radiative interface (see Eq.~\eqref{eqn:WBcond2}).\label{fig:IGW_dots}}
	\centering
\end{figure*} 
We calculated nonlinearity parameters using Eq.~\eqref{eq:propeqn} and Eq.~\eqref{eqn:WBcond2} for a chosen set of frequencies and wave numbers dictated by the generation spectra listed in Table~\ref{table:mandn} and present them in Fig. \ref{fig:IGW_damping}. The convective velocity ($u_c$), which is required to compute IGW generation wave amplitudes (see Section \ref{sec:IGW_gen}), was calculated by taking the average of all MESA-generated convective velocities in the grid points between the first and third quartile (in radius) of the convection zone, as shown below:
\begin{equation}
	u_c = u_{\mathrm{MLT}} = \frac{1}{\mathrm{N}_g/2}\sum_{\mathrm{N}_g/4}^{3\mathrm{N}_g/4} u_i
\end{equation}
where $i$ is grid space index inside the convection zone and $\mathrm{N}_g$ is the last grid index within the convection zone. The term $u_i$ refers to the convective velocity at the grid with index $i$. We label $u_c$ as $u_{\mathrm{MLT}}$, since the convective velocities in the convection zone are computed using mixing length theory (MLT). The convective turnover frequency, $\omega_c$, was set as
\begin{equation}\label{eq:ctf}
\omega_c = \frac{u_c}{r_c}.
\end{equation} 
In Fig.~\ref{fig:IGW_damping}, solid black horizontal lines in both plots represent $\epsilon = 1$ (strong nonlinearity condition) and $\epsilon = 0.1$ (weak nonlinearity condition). It is certainly the case that waves with $\epsilon > 1$ \citep{1981ApJ...245..286P} become nonlinear but nonlinear effects may also become important at lower $\epsilon$ \citep{2010MNRAS.404.1849B}. Note that from here on, our analysis of nonlinear waves (by either the strong or weak condition) will be only on those that become nonlinear beyond 0.5 $H_P$ from the convective-radiative interface, which is the nominal convective overshoot depth \citep{2013ApJ...772...21R}. We will refer to this as $r_{\mathrm{min,break}}$. 

Since the damping opacity is proportional to $\omega^{-4}$ and $\ell^{3}$ (see Eq.~\eqref{eq:gam_over_vg}), higher-frequency and lower-$\ell$ waves are expected to be damped less, without consideration of generation spectra. Thus, they are more likely to become nonlinear close to the stellar surface from amplitude growth alone. When all the generation spectra are considered as shown in Fig.~\ref{fig:IGW_damping}, we see that not only do the waves need to have higher frequency and lower wavenumber to become nonlinear close to the surface, they must also be generated with sufficient amplitudes. Nonlinear waves from spectrum F satisfy both these criteria for our example. On the other hand, low frequency waves that are generated with high amplitudes have a higher chance of becoming nonlinear close to the convective core. Figure~\ref{fig:IGW_damping} shows that as frequency decreases, the initial $\epsilon$ for all generation spectra increases. This means that before the effect of damping starts to dominate, these low-frequency waves can become  nonlinear near the core.

\subsection{IGW Generation Spectra Dependence}
The different line styles in Fig.~\ref{fig:IGW_damping} represent different IGW generation spectra introduced in Section~\ref{sec:IGW_gen}. It can be seen that when $l = 1$ (top panel), only 1 out of the 14 waves shown become nonlinear under the strong condition, which is the 10 $\mu$Hz wave generated from spectrum F. The flat spectrum allows high frequency waves to be generated with high amplitude. This, coupled with the fact that high-frequency waves experience lower damping, allows the 10 $\mu$Hz wave to become strongly nonlinear close to the stellar surface. Note that zero dependence on IGW frequency or wave number means that all waves generated from spectrum F start with the same initial amplitude. However, since the nonlinearity parameter is $k_h u_h/\omega$, the initial $\epsilon$ will be different for different frequencies and horizontal wave numbers. Waves from the other three generation spectra do not become strongly nonlinear, either because they are not generated with a large enough amplitude or because they experience large damping.

\begin{figure*}
	\centering
	\subfloat[Similar to Fig.~\ref{fig:IGW_dots} but with $u_c = u_{\mathrm{MLT}}/3$.]{
		\includegraphics[trim={0.0cm 0.0cm 0 0.0cm},clip,width=0.65\textwidth]{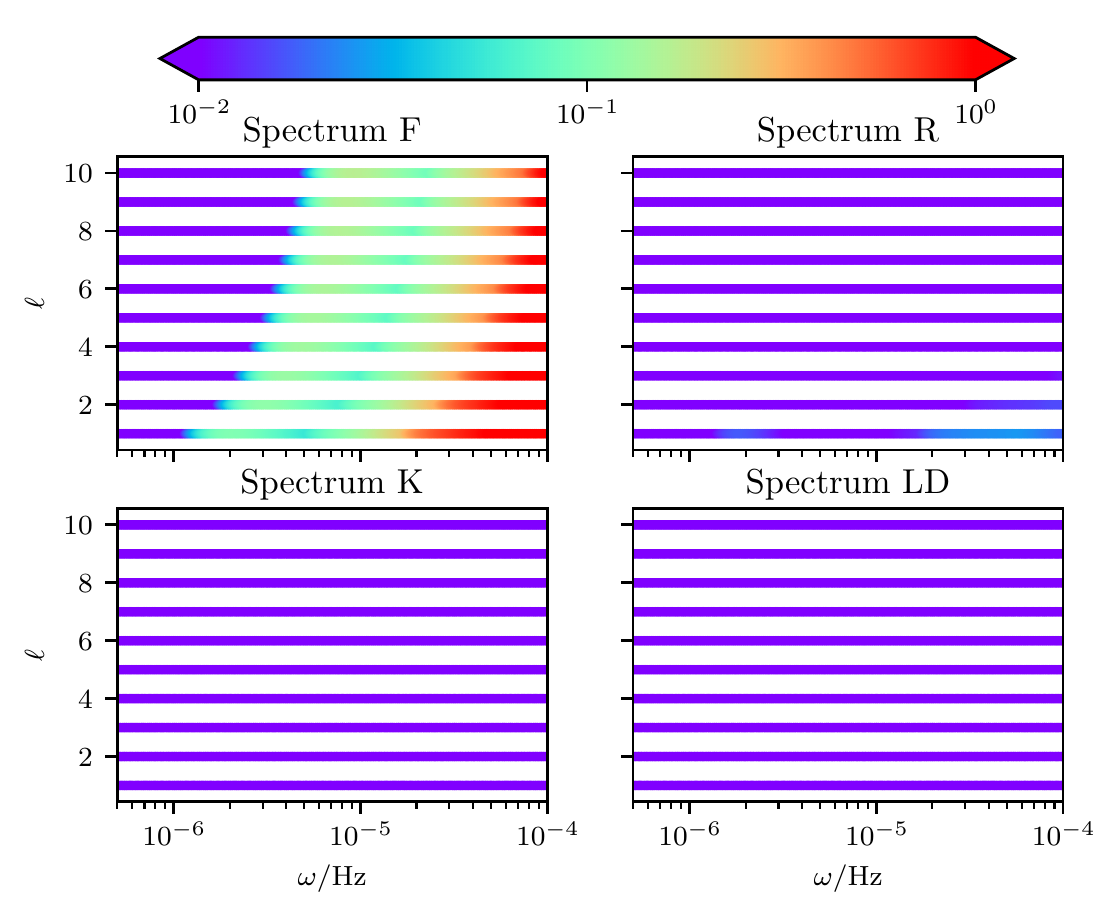}
		\label{fig:IGW_dots_1/3}}
	\qquad
	\subfloat[Similar to Fig.~\ref{fig:IGW_dots} but with $u_c = 3u_{\mathrm{MLT}}$.]{
		\includegraphics[trim={0.0cm 0.0cm 0 0.0cm},clip,width=0.65\textwidth]{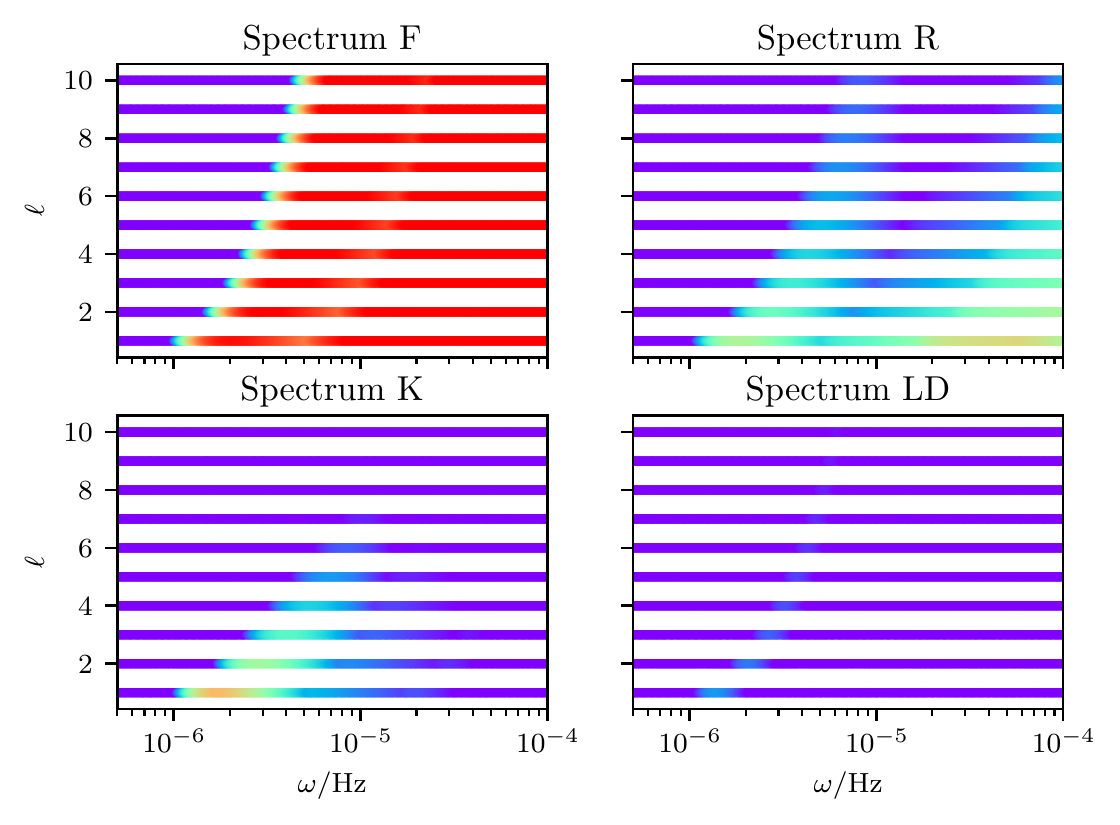}
		\label{fig:IGW_dots_3}}	
        \caption{Colour maps of nonlinear waves for different convective velocities. \label{fig:alldots}}
\end{figure*}

Looking at waves with $\epsilon > 0.1$ at $r > r_{\mathrm{min,break}}$, we can see one additional wave generated with spectrum F satisfying the weak condition. At lower frequencies, gravity waves are more likely to become nonlinear closer to the convective-radiative interface because of higher generation amplitudes and the boost due to the rapid increase of the \brunt{} frequency close to the interface. However, even though the initial amplitudes of low-frequency waves generated from spectrum K, spectrum LD and spectrum R are higher compared to their higher frequency counterparts, they are not boosted sufficiently to become nonlinear.

When $\ell$ is increased to 3 (bottom panel of Fig.~\ref{fig:IGW_damping}), a few differences can be noted in the $\epsilon$ behaviour compared to that for $\ell$ = 1. First, when waves of the same frequency are considered, small-scale waves are generated with higher $\epsilon$. Spectrum LD does this most efficiently, followed by spectrum K, spectrum F and spectrum R.  Second, increasing $\ell$ results in increased damping which causes the peaks of $\epsilon$ to be lower in the $l = 3$ case compared to the $l = 1$ case. This can lead to higher $\ell$ waves not achieving sufficient $\epsilon$ to become nonlinear. Thus, for a fixed frequency, the wave number determines which of the two effects mentioned above dominate.

Fig.~\ref{fig:IGW_dots} shows the nonlinearity parameter with a broader range of IGW horizontal wave numbers ($\ell$ = 1 -- 10) and frequencies ($\omega$ = 0.5 $\mu$Hz to 100 $\mu$Hz). The colour scheme represents the maximum $\epsilon$ reached by an IGW of a given wave number and frequency at $r > r_{\mathrm{min,break}}$. The range in frequency of 0.5 $\mu$Hz to 100 $\mu$Hz was chosen to look at the range of IGW thought to be generated by convection. The spacing between wave frequencies was chosen to be equal in the logarithmic scale. Note that in Fig.~\ref{fig:IGW_dots}, any nonlinearity parameter above 1.0 or below 0.01 will be red or blue respectively. The lower limit of $\epsilon$ has been set to 0.01 to provide a clear view of how $\epsilon$ varies across the weak condition. It was found that spectrum F has the highest fraction of nonlinear waves (for both strong and weak nonlinearity conditions), while spectrum K and spectrum LD show no nonlinear waves. Spectrum R shows that potentially nonlinear waves are concentrated in the lower $\ell$ regime. However, there is an irregular trend with increasing frequencies. At the lowest frequencies (0.5 $\mu$Hz - 1.0 $\mu$Hz), damping is at its highest causing most waves to be damped rapidly. Across all other tested frequencies at low-$\ell$, we see two peaks in $\epsilon$. The peak at low frequency is due to waves becoming nonlinear close to the convective core. The second peak at higher frequency is due to waves breaking at the surface. In between, waves are of high enough frequencies to not be generated with sufficient amplitude to break near the core but of low enough frequency to still be damped before reaching the surface. This damping is of course wavelength dependent, hence the structure seen in $\ell$. 

In conclusion, one can see that with MLT velocities, only waves from spectrum R and spectrum F can become nonlinear. 

\begin{figure*}
	\centering
	\includegraphics[trim={0.0cm 0.2cm 0 0.0cm},clip,width=0.75\textwidth]{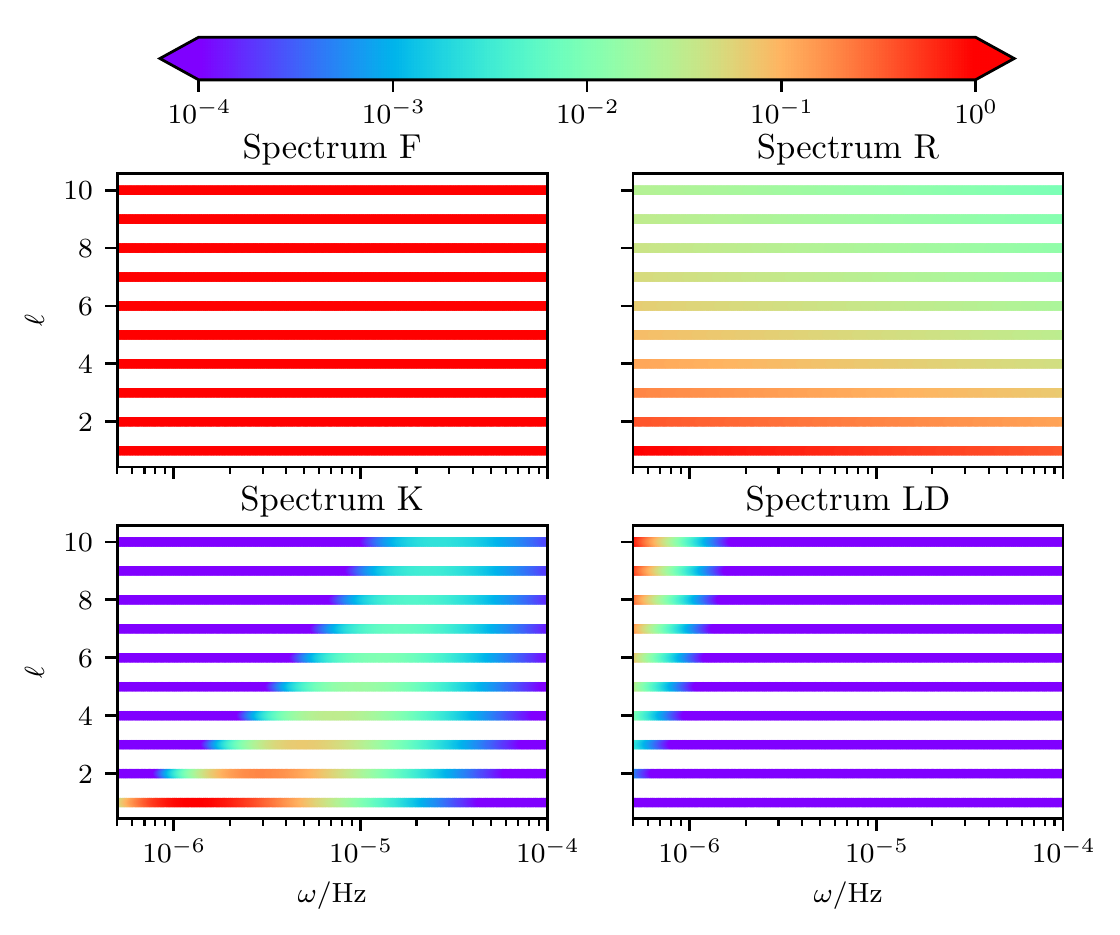}
	\caption{Colour maps of the nonlinear wave energies normalised to the maximum wave energy for different spectra at $u_c = 3\;u_{\mathrm{MLT}}$. At $u_c = \;u_{\mathrm{MLT}}$, the plots show similar trends to the case of $u_c = 3\;u_{\mathrm{MLT}}$, with the exception of spectrum K, a smaller spread of high wave energies over frequency for higher $\ell$ and lower $\omega$ waves. \label{fig:IGW_dots_energy_uc3}}
	\centering
\end{figure*}

\begin{figure}
	\centering
	\includegraphics[trim={0.0cm 0.0cm 0 0.0cm},clip,width=1.0\columnwidth]{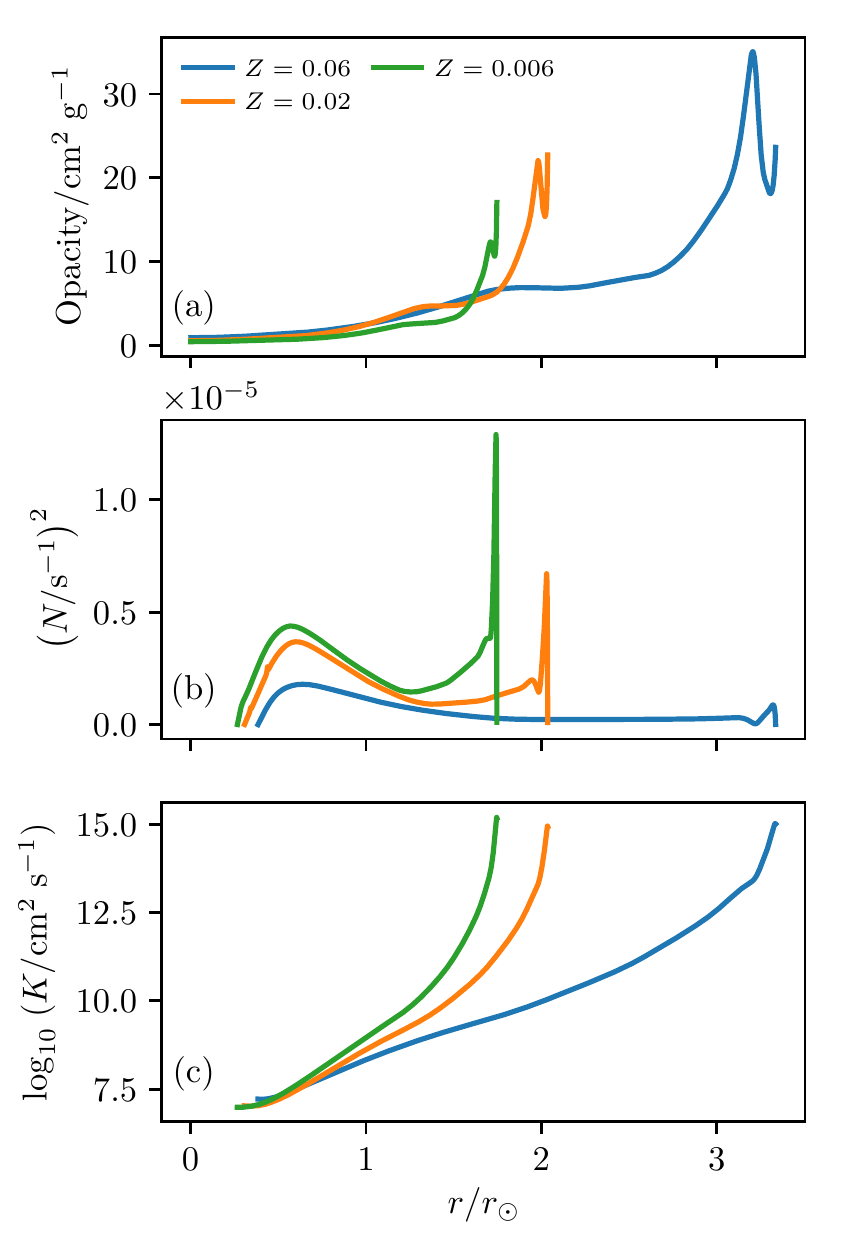}
	\vspace{-20pt}
	\caption{(a) Opacity, (b) \brunt{} frequency squared and the (c) thermal diffusivity profiles as a function of radius, in units of solar radius. Colour representations for all three plots are stated in the legend of the top panel. The opacity profiles (top panel) for all three metallicities show bumps occurring at radii where $\log T = 5.3$, which is known as the iron opacity bump. This bump, together with temperature and pressure profiles are responsible for the bump seen in the \brunt{} frequency squared profile (middle panel). Note that higher metallicities lead to shallower thermal diffusivity slopes but the surface $K$ values are approximately equal (bottom panel).  \label{fig:diff_Z}}
	\centering	
\end{figure}

\subsection{Convective Velocity Dependence}\label{sec:conv_vel}
The bulk convective velocities, $u_c$, were obtained from the formulation of mixing length theory \citep{1951ZA.....29..274B,1958ZA.....46..108B} in MESA. While being widely adopted, it is also well known that this prescription is likely not an accurate representation of the three-dimensional nature of turbulent convection. In particular, simulations of turbulent convection generally show significantly more intermittency than this prescription allows for. In addition to the uncertainty related to MLT, the MLT velocity itself, $u_{\mathrm{MLT}}$, can vary significantly within the convection zone. Generally, within the convection zone of a MESA model, the MLT velocity varied by approximately one order of magnitude for all the masses investigated. Furthermore, a recent work \citep{CoustonLouis-Alexandre2018Tefs}, which looked at 3D simulations of IGW generated by turbulent convection, has shown that at the point of generation, the steep fall-off in frequency dependence as in spectrum K and spectrum LD occurs well away from the traditional convective turnover frequency relation (a factor of 2-5). This introduces further uncertainties to a factor of 2-5 in the value of $u_c$. Thus, we investigated the effect of varying convective velocities by varying $u_c$ from one-third to three times $u_{\mathrm{MLT}}$.
Fig. \ref{fig:IGW_dots_1/3} and Fig. \ref{fig:IGW_dots_3} show how the nonlinearity parameter changes with higher and lower convective velocities. In Fig. \ref{fig:IGW_dots_1/3}, where $u_c$ was set to one-third of the initial convective velocity given by MESA, only waves from spectrum F show strong nonlinearity. On the other hand, when $u_c$ was set to 3 $u_{\mathrm{MLT}}$, waves from all spectra, except from spectrum LD, exhibit strong nonlinearity

For a larger $u_c$, spectrum K can be seen to exhibit less nonlinear waves at high frequencies as these waves are generated with lower amplitudes. At low frequencies and larger $\ell$, generation amplitudes are larger but damping is also larger. Moreover, the exponential term reduces amplitudes further. This causes only a small range of waves with specific frequencies to become nonlinear. Whether waves become nonlinear depends sensitively on the amplitudes of convective velocity which may very well be time-dependent. Hence, it is entirely possible that the occurence of breaking waves is also time-dependent.

\subsection{IGW Energies}\label{sec:IGW_energy}
To better understand the impact breaking waves would have in stellar processes like mixing and angular momentum transfer, we look at the energy in waves that become nonlinear, which can be represented by the following relation:
\begin{equation*}
E \propto u_h^2 
\end{equation*}
For the flat spectrum, the nonlinear wave energies are distributed equally for all wave numbers and frequencies. For spectrum R, 
\begin{align*}
E &\propto \left(\sqrt{\ell(\ell + 1)}\right)^{-1.8}\omega^{-1.2},
\end{align*}
which means that waves with the lowest wave numbers and lowest frequencies possess the highest energies. On the other hand, for spectrum LD,
\begin{equation*}
E \propto \left(\sqrt{\ell(\ell + 1)}\right)^4 \omega^{-8.50}.
\end{equation*}
Therefore, the waves that carry the most energy are the high-wave number, low-frequency waves. Finally, for spectrum K,
\begin{equation*}
E \propto \left(\sqrt{\ell(\ell + 1)}\right)^2 \omega^{-4.34} \exp \left[- k_h^2 \left( \frac{u_c}{\omega_c}\right)^2\left( \frac{\omega}{\omega_c}\right)^{-4/3} \right].
\end{equation*}
The distribution of wave energies for spectrum K is not as direct as for the other spectra due to the exponential term. In Figure~\ref{fig:IGW_dots_energy_uc3}, which shows the nonlinear wave energy trend for all generation spectra, we can see that for spectrum K and $u_c = u_{\mathrm{MLT}}$, the nonlinear wave energies are highest for low-$\ell$ and low-$\omega$ waves.  

The ratio of energy in waves that satisfy $\epsilon \geqslant 0.1$ to the total energy of all waves within the frequency and wave number ranges considered, is
\begin{equation}\label{eq:energy_ratio}
\frac{E_{\epsilon \geqslant 0.1}}{E_{\mathrm{Total}}} = \frac{\left(\sum_{l,\omega} u_h^2 \;\Delta\omega \right)_{\epsilon \geqslant 0.1}}{\sum_{l,\omega} u_h^2 \; \Delta\omega},
\end{equation}
where the numerator on the right-hand side represents the sum over IGW amplitudes that satisfy $\epsilon \geqslant 0.1$, at radii larger than $r_{\mathrm{min,break}}$, while the denominator is the total energy input in waves. The spacing between frequencies is represented by $\Delta\omega$. The spacing between $\ell$ is 1. Both energy terms in the numerator and the denominator are energies supplied by convective processes, so this ratio was calculated at the point of wave generation.

Applying Eq.~\eqref{eq:energy_ratio} to the same stellar model for cases of $u_c = u_{\mathrm{MLT}}$ and $u_c = 3\;u_{\mathrm{MLT}}$ gives the values shown in Table~\ref{table:perc_metal} at $Z = 0.02$. Energy ratios for spectrum F are always close to one as most waves generated from spectrum F become nonlinear as shown in Fig.~\ref{fig:IGW_dots} and Fig.~\ref{fig:alldots}. For $u_c = u_{\mathrm{MLT}}$, only spectrum F shows any appreciable energy in nonlinear waves (see Fig.~\ref{fig:IGW_dots}). When $u_c$ is increased to $3\;u_{\mathrm{MLT}}$, all the generation spectra, except spectrum LD, produce nonlinear waves. Nonlinear waves from spectrum R and spectrum K have moderately higher energy ratios (see Table~\ref{table:perc_metal}). This due to the significant overlaps between the nonlinearity parameter trend (see Fig.~\ref{fig:IGW_dots_3}) and the wave energy trend (Fig.~\ref{fig:IGW_dots_energy_uc3}), for both spectrum R and spectrum K. 

\subsection{Metallicity Dependence}\label{sec:results_metal}
In the results discussed above, we have used a stellar model generated with an initial metallicity of 0.02, equal to the solar metallicity. To investigate the effect of different initial metallicities on the production of nonlinear waves, we generated stellar models at ZAMS with initial metallicities of 0.006 and 0.06, and tested these models with different wave generation spectra \text{at $u_c$ = $u_{\mathrm{MLT}}$ and $u_c$ = 3 $u_{\mathrm{MLT}}$}. As in the previous section, we have chosen a fixed nonlinearity condition of $\epsilon \geqslant 0.1$. We calculate energy ratios based on Eq.~\eqref{eq:energy_ratio} and show them in Table~\ref{table:perc_metal}.

\begin{table}
	\centering		
	\begin{tabular}{ccccc}
		\hline \hline
		Spectra & $u_c/u_{\mathrm{MLT}}$ & $Z$ = 0.06 & $Z$ = 0.02 & $Z$ = 0.006  \\ \hline \hline 
        F & 1 & 0.968 & 0.968 & 0.959 \\
         & 3  & 0.971 & 0.9705 & 0.9635 \\ \hline
		R & 1 & 0.0 & 0.0 & $5.54 \times 10^{-2}$ \\
         & 3  & 0.234 & 0.223 & 0.182 \\ \hline
        K & 1  & 0.0 & 0.0 & 0.0 \\
		  & 3  & 0.481 & 0.480 & 0.606 \\ \hline
        LD & 1  & 0.0 & 0.0 & 0.0 \\
		   & 3  & 0.0 & 0.0 & 0.0 \\ \hline
	\end{tabular}
	\caption{The table shows the energy ratios of nonlinear waves for different stellar metallicities at ZAMS.}
	\label{table:perc_metal}
\end{table}

The trend with spectrum F shows a slight decrease with decreasing metallicities due to increasing damping as lower metallicties lead to lower opacities (see Fig~\ref{fig:diff_Z}\textcolor{blue}{a}). Other than spectrum F, virtually no other spectrum shows nonlinear waves for any metallicity at $u_c$ = $u_{\mathrm{MLT}}$. The exception is spectrum R which shows a slight increase in nonlinear wave energy at $Z$ = 0.006 because the product of the density stratification term and geometric term (see Fig.~\ref{fig:IGW_propagation}) becomes larger due to decreasing total stellar radius (as shown in Fig.~\ref{fig:diff_Z}) and increasing \brunt{} frequency spike (see Fig.~\ref{fig:diff_Z}\textcolor{blue}{b}) causing wave amplification close to the stellar surface.

Moving to $u_c$ = 3 $u_{\mathrm{MLT}}$, we see that spectrum R and spectrum K have opposite trends in nonlinear energy ratios for decreasing metallicity. Spectrum R shows a decreasing trend because of the same reason spectrum F shows a decreasing trend. Spectrum K shows an increasing trend as convective velocities are found to be higher when stellar metallicity is lower.

\subsection{Age and Mass Dependences}\label{sec:AgeMass}
To investigate whether there are any trends of waves becoming nonlinear with varying stellar mass and age, we extended our analysis to include stellar masses up to 20 $M_{\odot}$ and $X_c$ down to 0.01, where we have used the central hydrogen mass fraction, $X_c$, as a proxy for stellar age. A fixed nonlinearity condition of $\epsilon \geqslant 0.1$ was chosen. 

\begin{figure}
	\centering
	\includegraphics[trim={0.0cm 0.0cm 0 0.0cm},clip,width=1.0\columnwidth]{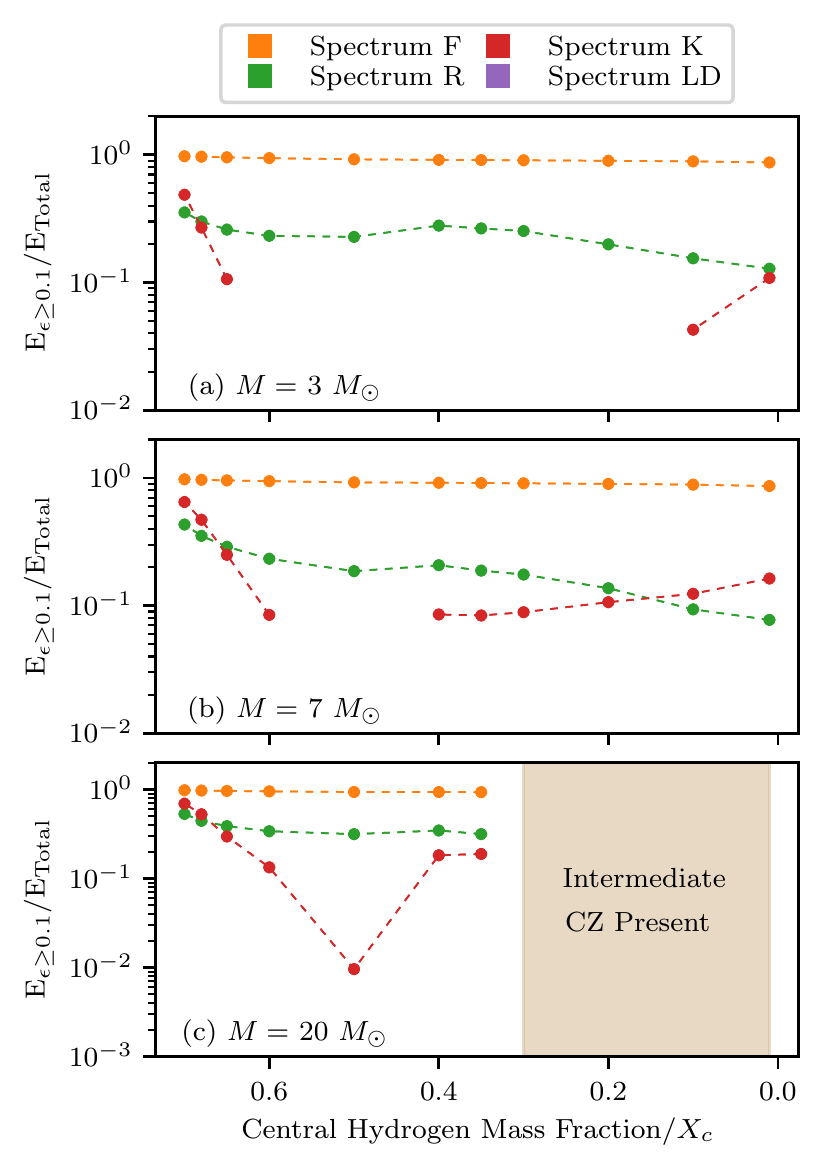}
	\vspace{-20pt}
	\caption{Energy ratios of waves with $\epsilon \geqslant 0.1$ to total energy from all waves against central hydrogen mass fractions for stars with masses (a) 3 $M_{\odot}$ , (b) 7 $M_{\odot}$ and (c) 20 $M_{\odot}$. The dotted lines with circular markers represent waves with $u_c = 3\; u_{\mathrm{MLT}}$. The highlighted region in the plot for 20 $M_{\odot}$ shows where the intermediate convection zone (CZ) starts to develop. The missing data points for spectrum K is due to no waves becoming nonlinear with our chosen nonlinearity parameter limit.  \label{fig:diff_age}}
	\centering	
\end{figure}

Figure \ref{fig:diff_age} shows how these energy ratios change as a function of $X_c$, with $u_c = $ 3 $u_{\mathrm{MLT}}$, for 3 $M_{\odot}$, 7 $M_{\odot}$ and 20 $M_{\odot}$ main-sequence stars. From midMS onwards, no data is plotted for 20 $M_{\odot}$ (bottom panel of Fig.~\ref{fig:diff_age}), as the star develops a convection zone in the radiation zone between the convective core and the surface convection zone. A quick analysis on waves generated from the intermediate convection zone showed that no waves become nonlinear for all spectra except spectrum F. This was due to the very small convective velocities generated in this narrow convection zone. Waves generated with such small amplitude never reach sufficient amplitudes to become nonlinear. This intermediate zone also significantly attenuates waves from the core.

\begin{figure}
	\centering
	\includegraphics[trim={0.0cm 0.0cm 0 0.0cm},clip,width=1.0\columnwidth]{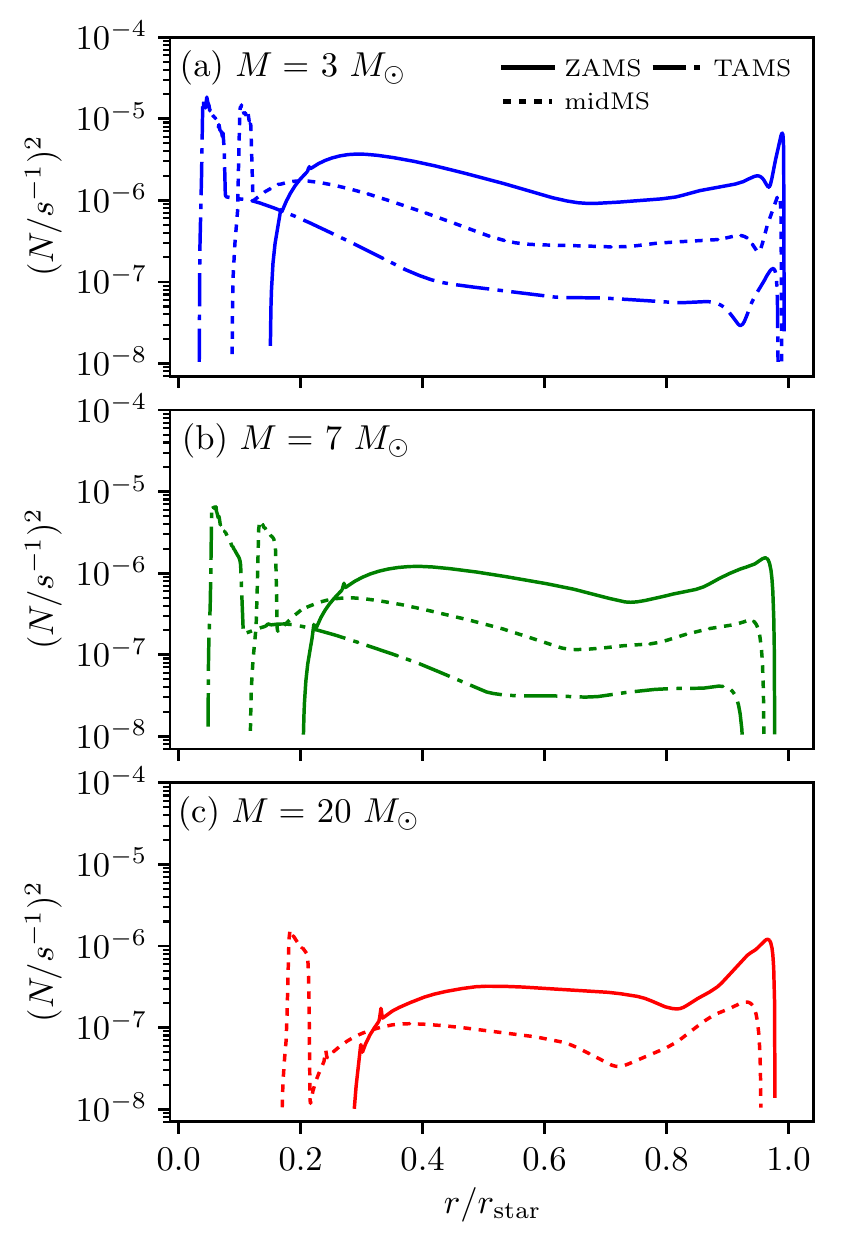}
	\vspace{-20pt}
	\caption{The \brunt{} frequency profile in the radiation zone for 8 stellar models. The top panel represents the profiles for a (a) 3 $M_{\odot}$ star, the middle panel for a (b) 7 $M_{\odot}$ star and the bottom panel for a (c) 20 $M_{\odot}$ star. The x-axis represents stellar radius in units of total stellar radius. \label{fig:brunt}}
	\centering	
\end{figure}

\begin{figure}
	\centering
	\includegraphics[trim={0.0cm 0.0cm 0 0.0cm},clip,width=1.0\columnwidth]{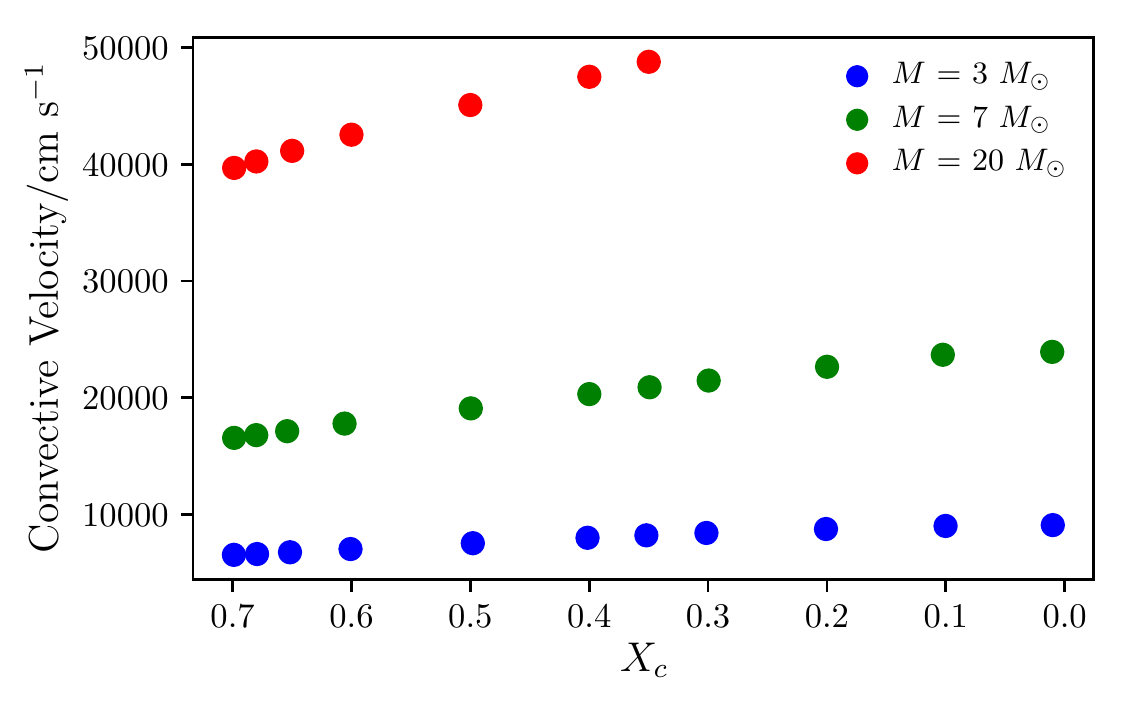}
	\vspace{-20pt}
	\caption{The bulk convective velocities, $u_{\mathrm{MLT}}$ as a function of central hydrogen mass fraction, $X_c$. Refer to Fig.~\ref{fig:brunt} for colour representations. \label{fig:cv}}
	\centering	
\end{figure}

In general, we see a decreasing trend of nonlinear wave energies with increasing age for spectrum F, spectrum R and initially for spectrum K. In fact, for spectrum K, this decrease leads to no waves becoming nonlinear with the effect being more prominent at lower stellar masses. To understand this, we look at various factors that affect the wave amplitude. First, Fig.~\ref{fig:brunt} shows that when a star starts to undergo hydrogen burning and evolve from ZAMS, its core contracts slowly leaving behind a steep hydrogen abundance gradient \citep{1991MNRAS.249..602D}. This leads to the formation of a sharp ``spike" in the \brunt{} frequency profile near the edge of convective core, which broadens and moves inwards with the convective core surface (see Fig.~\ref{fig:brunt}). This spike effectively traps low-frequency, high-wave number waves causing less waves to become nonlinear close to the stellar surface. Moreover, the \brunt{} frequencies are lower for older stars which means the density stratification term (from Eq.~\eqref{eq:propeqn}) will be smaller too. Furthermore, as the star ages, stellar density varies over the same magnitude across the radiation zone but the star grows in size. This causes the density gradient to decrease and the geometric term to grow, leading to their product (as shown in Eq.~\eqref{eq:propeqn}) being smaller in older stars. It was thus observed that despite the increase in convective velocities with age (see Fig.~\ref{fig:cv}), which should lead to an increasing trend of nonlinear wave energies, the trend seen in Fig.~\ref{fig:diff_age} shows that the factors mentioned above dominate. 

We see an increasing trend in nonlinear wave energy ratios for spectrum K at higher ages. To explain this, we look at Fig.~\ref{fig:brunt}, which shows that the peak \brunt{} frequency close to the stellar surface is lower for older stars. This causes higher frequency waves to experience lower damping close to the surface which allows more waves to cross the $\epsilon > 0.1$ threshold.

\begin{figure}
	\centering
	\includegraphics[trim={0.0cm 0.0cm 0 0.0cm},clip,width=1.0\columnwidth]{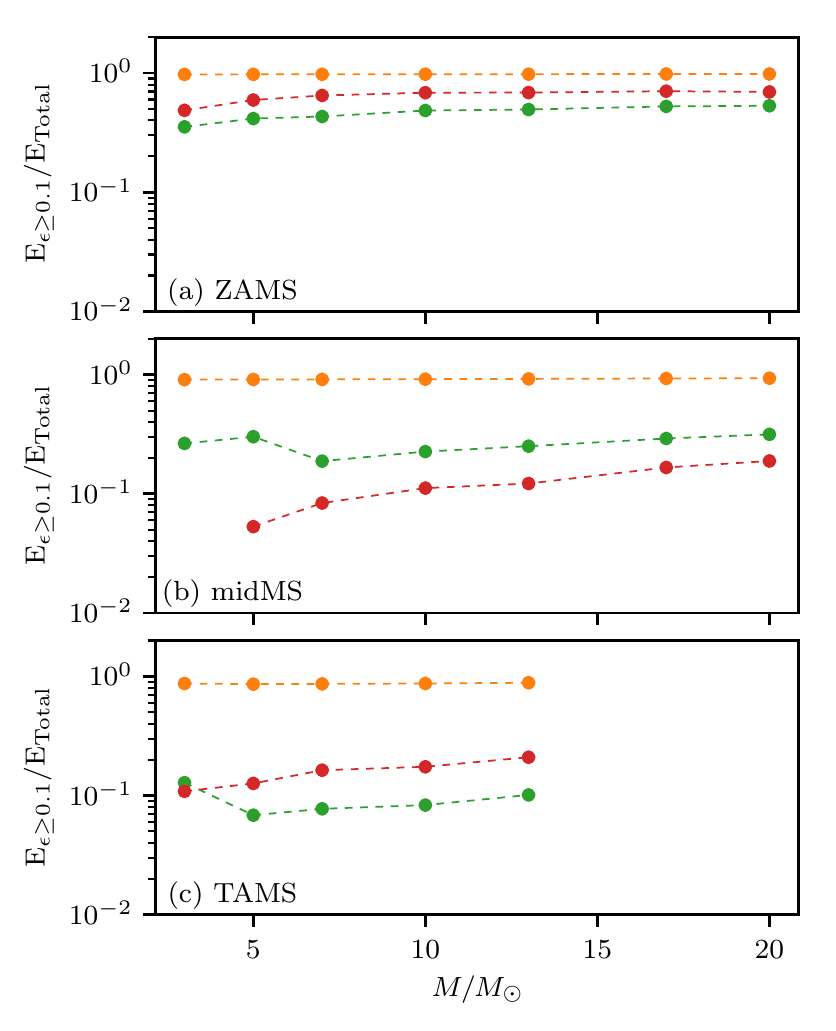}
	\vspace{-20pt}
	\caption{Energy ratios of waves against stellar masses at (a) ZAMS, (b) midMS and (c) TAMS. Refer to Fig.~\ref{fig:diff_age} for the colour representation. Note that we have omitted the 17 $M_{\odot}$ energy ratio here as the \brunt{} frequencies in most of the radiation zone were found to be close to or less than 100 $\mu$Hz.  \label{fig:diff_mass}}
	\centering
\end{figure}    

Figure \ref{fig:diff_mass} shows wave energy ratios as a function of mass with the top panel representing ZAMS models, the middle panel representing midMs models and the bottom panel representing TAMS models. As in the case for age dependence, energy ratios of waves from spectrum R are higher than energy ratios of waves from spectrum K. In general, the nonlinear energy ratios for all spectra increase slightly with increasing mass. This can be explained by the increasing convective velocities with mass (see Fig.~\ref{fig:cv}). Furthermore, Fig.~\ref{fig:brunt} shows that for larger stellar masses, the \brunt{} frequencies are lower, which leads to lower damping.

\section{Discussion and Conclusion}\label{sec:dis}

In this paper, we have shown that depending on the spectrum and amplitudes of waves generated by convection, some waves may become nonlinear in the radiative regions of intermediate-mass stars. This can occur because the amplitudes of high(er) frequency/longer wavelength waves are amplified by density stratification. Therefore, if such waves are generated with sufficient amplitude, they can become nonlinear along their journey towards the stellar surface. On the other hand, lower frequency waves, which are generated with larger amplitudes, can become nonlinear close to the convective core. These waves that break near the convective core may lead to enhanced mixing beyond a nominal ``overshoot depth''. 

One crucial element as to whether waves can become nonlinear and break in stellar radiative regions is the wave generation spectrum. Compared to the theoretical generation spectra from \cite{2013MNRAS.430.2363L} and \cite{1999ApJ...520..859K}, with their steep fall off at frequencies higher than the convective turnover frequency, the numerical generation spectra from \cite{2013ApJ...772...21R} show that more waves may become nonlinear and consequently, more energy is in these nonlinear waves for MLT convective velocities. 

The assumed convective velocity also plays an important role in the ability of waves to become nonlinear, with even a factor of three increase causing significant differences in total nonlinear wave energies. One such significant difference occurs when convective velocities are set as three times the MLT values. In this case, nonlinear wave energy ratios from the theoretical generation spectrum of \cite{1999ApJ...520..859K} became comparable to those from the numerical generation spectrum R. Given the likelihood of variation in convective velocities by at least this order, it is worth trying to understand the role that intermittency plays in determining the wave spectrum. This is important especially after the results from a recent numerical work from \cite{CoustonLouis-Alexandre2018Tefs} which shows that the steep fall-off in energy with frequency dependence occurs away from the classical theory for convective turnover frequency. Indeed, it may be possible that the wave generating process at the convective-radiative interface and wave breaking process at the surface may be time-dependent.

We found that age, mass and metallicity also affect the fraction of waves becoming nonlinear, but to a lesser degree.  Generally, as stars become older the fraction of energy that goes into nonlinear waves decreases because the receding convective core leaves behind a compositional gradient and hence, spike in the \brunt{} frequency which effectively filters waves from propagating outward.  As the mass increases, the fraction of nonlinear waves increases slightly due to increasing convective velocities. As the metallicity increases, the fraction of nonlinear waves behaves differently, depending on generation spectra.  

Whether or not waves can become nonlinear in the radiative regions of stars plays an important role on the efficiency with which waves can transport momentum and mix chemical elements. Clearly, wave breaking and the ensuing turbulence that is generated would lead to more efficient mixing and angular momentum transport than simple radiative dissipation. The strong observational evidence for enhanced coupling between convective and radiative regions \citep{2012Natur.481...55B,0004-637X-756-1-19,2013ApJ...775L...1T} and the possible observational detections of IGW\citep{2015ApJ...806L..33A,2017A&A...602A..32A,2018MNRAS.476.1234A,2018MNRAS.480..972R} both point to breaking IGW, which in turn, points to a flatter frequency generation spectrum from convection than what theoretical models predict or higher convective velocities than those predicted by MLT. It is therefore vital that more theoretical and numerical investigations into the convective generation spectrum of IGW, and the physics that drives it, are undertaken. In particular, it is worth further studying plume models, which depending on parameter choices, can produce generation spectra with flatter dependences on frequencies \citep{pincon2016a}.

\section*{Acknowledgments}
We acknowledge support from STFC grant ST/L005549/1 and NASA grant NNX17AB92G. Computing was carried out on Pleiades at NASA Ames. We thank Conny Aerts, Daniel Lecoanet, May Gade Pedersen and Dominic Bowman for useful conversations leading to the development of this manuscript. Finally, we would like to thank Manoj Previn Ratnasingam and Romy Klaasje Verhoeven for the proofreading support they provided. 




\bibliographystyle{mnras}
\bibliography{bi} 

\newcommand{\noop}[1]{}
\begin{thebibliography}{}
\makeatletter
\relax
\def\mn@urlcharsother{\let\do\@makeother \do\$\do\&\do\#\do\^\do\_\do\%\do\~}
\def\mn@doi{\begingroup\mn@urlcharsother \@ifnextchar [ {\mn@doi@}
  {\mn@doi@[]}}
\def\mn@doi@[#1]#2{\def\@tempa{#1}\ifx\@tempa\@empty \href
  {http://dx.doi.org/#2} {doi:#2}\else \href {http://dx.doi.org/#2} {#1}\fi
  \endgroup}
\def\mn@eprint#1#2{\mn@eprint@#1:#2::\@nil}
\def\mn@eprint@arXiv#1{\href {http://arxiv.org/abs/#1} {{\tt arXiv:#1}}}
\def\mn@eprint@dblp#1{\href {http://dblp.uni-trier.de/rec/bibtex/#1.xml}
  {dblp:#1}}
\def\mn@eprint@#1:#2:#3:#4\@nil{\def\@tempa {#1}\def\@tempb {#2}\def\@tempc
  {#3}\ifx \@tempc \@empty \let \@tempc \@tempb \let \@tempb \@tempa \fi \ifx
  \@tempb \@empty \def\@tempb {arXiv}\fi \@ifundefined
  {mn@eprint@\@tempb}{\@tempb:\@tempc}{\expandafter \expandafter \csname
  mn@eprint@\@tempb\endcsname \expandafter{\@tempc}}}

\bibitem[\protect\citeauthoryear{{Aerts} \& {Rogers}}{{Aerts} \&
  {Rogers}}{2015}]{2015ApJ...806L..33A}
{Aerts} C.,  {Rogers} T.~M.,  2015, \mn@doi [\apjl]
  {10.1088/2041-8205/806/2/L33}, \href
  {http://ukads.nottingham.ac.uk/abs/2015ApJ...806L..33A} {806, L33}

\bibitem[\protect\citeauthoryear{{Aerts}, {Puls}, {Godart}  \&
  {Dupret}}{{Aerts} et~al.}{2009}]{2009A&A...508..409A}
{Aerts} C.,  {Puls} J.,  {Godart} M.,   {Dupret} M.-A.,  2009, \mn@doi [\aap]
  {10.1051/0004-6361/200810471}, \href
  {http://ukads.nottingham.ac.uk/abs/2009A%26A...508..409A} {508, 409}

\bibitem[\protect\citeauthoryear{Aerts, Christensen-Dalsgaard  \& Kurtz}{Aerts
  et~al.}{2010}]{aerts2010asteroseismology}
Aerts C.,  Christensen-Dalsgaard J.,   Kurtz D.,  2010, Asteroseismology.
Astronomy and Astrophysics Library, Springer Netherlands, \url
  {https://books.google.co.uk/books?id=N8pswDrdSyUC}

\bibitem[\protect\citeauthoryear{{Aerts} et~al.,}{{Aerts}
  et~al.}{2017}]{2017A&A...602A..32A}
{Aerts} C.,  et~al., 2017, \mn@doi [\aap] {10.1051/0004-6361/201730571}, \href
  {http://adsabs.harvard.edu/abs/2017A%26A...602A..32A} {602, A32}

\bibitem[\protect\citeauthoryear{{Aerts} et~al.,}{{Aerts}
  et~al.}{2018}]{2018MNRAS.476.1234A}
{Aerts} C.,  et~al., 2018, \mn@doi [\mnras] {10.1093/mnras/sty308}, \href
  {http://adsabs.harvard.edu/abs/2018MNRAS.476.1234A} {476, 1234}

\bibitem[\protect\citeauthoryear{{Alvan}, {Brun}  \& {Mathis}}{{Alvan}
  et~al.}{2014}]{2014A&A...565A..42A}
{Alvan} L.,  {Brun} A.~S.,   {Mathis} S.,  2014, \mn@doi [\aap]
  {10.1051/0004-6361/201323253}, \href
  {http://ukads.nottingham.ac.uk/abs/2014A%26A...565A..42A} {565, A42}

\bibitem[\protect\citeauthoryear{{Alvan}, {Strugarek}, {Brun}, {Mathis}  \&
  {Garcia}}{{Alvan} et~al.}{2015}]{2015A&A...581A.112A}
{Alvan} L.,  {Strugarek} A.,  {Brun} A.~S.,  {Mathis} S.,   {Garcia} R.~A.,
  2015, \mn@doi [\aap] {10.1051/0004-6361/201526250}, \href
  {http://ukads.nottingham.ac.uk/abs/2015A%26A...581A.112A} {581, A112}

\bibitem[\protect\citeauthoryear{{Ansong} \& {Sutherland}}{{Ansong} \&
  {Sutherland}}{2010}]{ansong_sutherland_2010}
{Ansong} J.~K.,  {Sutherland} B.~R.,  2010, \mn@doi [Journal of Fluid
  Mechanics] {10.1017/S0022112009993193}, 648, 405–434

\bibitem[\protect\citeauthoryear{{Barker} \& {Ogilvie}}{{Barker} \&
  {Ogilvie}}{2010}]{2010MNRAS.404.1849B}
{Barker} A.~J.,  {Ogilvie} G.~I.,  2010, \mn@doi [\mnras]
  {10.1111/j.1365-2966.2010.16400.x}, \href
  {http://ukads.nottingham.ac.uk/abs/2010MNRAS.404.1849B} {404, 1849}

\bibitem[\protect\citeauthoryear{{Beck} et~al.,}{{Beck}
  et~al.}{2012}]{2012Natur.481...55B}
{Beck} P.~G.,  et~al., 2012, \mn@doi [\nat] {10.1038/nature10612}, \href
  {http://adsabs.harvard.edu/abs/2012Natur.481...55B} {481, 55}

\bibitem[\protect\citeauthoryear{{Biermann}}{{Biermann}}{1951}]{1951ZA.....29..274B}
{Biermann} L.,  1951, \zap, \href
  {http://adsabs.harvard.edu/abs/1951ZA.....29..274B} {29, 274}

\bibitem[\protect\citeauthoryear{{B{\"o}hm-Vitense}}{{B{\"o}hm-Vitense}}{1958}]{1958ZA.....46..108B}
{B{\"o}hm-Vitense} E.,  1958, \zap, \href
  {http://adsabs.harvard.edu/abs/1958ZA.....46..108B} {46, 108}

\bibitem[\protect\citeauthoryear{{Brummell}, {Clune}  \& {Toomre}}{{Brummell}
  et~al.}{2002}]{2002ApJ...570..825B}
{Brummell} N.~H.,  {Clune} T.~L.,   {Toomre} J.,  2002, \mn@doi [\apj]
  {10.1086/339626}, \href
  {http://ukads.nottingham.ac.uk/abs/2002ApJ...570..825B} {570, 825}

\bibitem[\protect\citeauthoryear{{Brun}, {Miesch}  \& {Toomre}}{{Brun}
  et~al.}{2011}]{2011ApJ...742...79B}
{Brun} A.~S.,  {Miesch} M.~S.,   {Toomre} J.,  2011, \mn@doi [\apj]
  {10.1088/0004-637X/742/2/79}, \href
  {http://ukads.nottingham.ac.uk/abs/2011ApJ...742...79B} {742, 79}

\bibitem[\protect\citeauthoryear{{Charbonnel} \& {Talon}}{{Charbonnel} \&
  {Talon}}{2005}]{2005Sci...309.2189C}
{Charbonnel} C.,  {Talon} S.,  2005, \mn@doi [Science]
  {10.1126/science.1116849}, \href
  {http://ukads.nottingham.ac.uk/abs/2005Sci...309.2189C} {309, 2189}

\bibitem[\protect\citeauthoryear{Couston, Lecoanet, Favier  \& Le~Bars}{Couston
  et~al.}{2018}]{CoustonLouis-Alexandre2018Tefs}
Couston L.-A.,  Lecoanet D.,  Favier B.,   Le~Bars M.,  2018, Journal of fluid
  mechanics, 854

\bibitem[\protect\citeauthoryear{Deheuvels et~al.,}{Deheuvels
  et~al.}{2012}]{0004-637X-756-1-19}
Deheuvels S.,  et~al., 2012, The Astrophysical Journal, 756, 19

\bibitem[\protect\citeauthoryear{{Denissenkov} \& {Tout}}{{Denissenkov} \&
  {Tout}}{2003}]{2003MNRAS.340..722D}
{Denissenkov} P.~A.,  {Tout} C.~A.,  2003, \mn@doi [\mnras]
  {10.1046/j.1365-8711.2003.06284.x}, \href
  {http://adsabs.harvard.edu/abs/2003MNRAS.340..722D} {340, 722}

\bibitem[\protect\citeauthoryear{{Dintrans}, {Brandenburg}, {Nordlund}  \&
  {Stein}}{{Dintrans} et~al.}{2003}]{2003Ap&SS.284..237D}
{Dintrans} B.,  {Brandenburg} A.,  {Nordlund} {\AA}.,   {Stein} R.~F.,  2003,
  \apss, \href {http://ukads.nottingham.ac.uk/abs/2003Ap%26SS.284..237D} {284,
  237}

\bibitem[\protect\citeauthoryear{{Dziembowski}, {Pamiatnykh}  \&
  {Sienkiewicz}}{{Dziembowski} et~al.}{1991}]{1991MNRAS.249..602D}
{Dziembowski} W.~A.,  {Pamiatnykh} A.~A.,   {Sienkiewicz} R.~.,  1991, \mn@doi
  [\mnras] {10.1093/mnras/249.4.602}, \href
  {http://adsabs.harvard.edu/abs/1991MNRAS.249..602D} {249, 602}

\bibitem[\protect\citeauthoryear{{Fritts} \& {Alexander}}{{Fritts} \&
  {Alexander}}{2003}]{2003RvGeo..41.1003F}
{Fritts} D.~C.,  {Alexander} M.~J.,  2003, \mn@doi [Reviews of Geophysics]
  {10.1029/2001RG000106}, \href
  {http://ukads.nottingham.ac.uk/abs/2003RvGeo..41.1003F} {41, 3}

\bibitem[\protect\citeauthoryear{{Fuller}, {Lecoanet}, {Cantiello}  \&
  {Brown}}{{Fuller} et~al.}{2014}]{2014ApJ...796...17F}
{Fuller} J.,  {Lecoanet} D.,  {Cantiello} M.,   {Brown} B.,  2014, \mn@doi
  [\apj] {10.1088/0004-637X/796/1/17}, \href
  {http://ukads.nottingham.ac.uk/abs/2014ApJ...796...17F} {796, 17}

\bibitem[\protect\citeauthoryear{{Fuller}, {Cantiello}, {Lecoanet}  \&
  {Quataert}}{{Fuller} et~al.}{2015}]{2015ApJ...810..101F}
{Fuller} J.,  {Cantiello} M.,  {Lecoanet} D.,   {Quataert} E.,  2015, \mn@doi
  [\apj] {10.1088/0004-637X/810/2/101}, \href
  {http://ukads.nottingham.ac.uk/abs/2015ApJ...810..101F} {810, 101}

\bibitem[\protect\citeauthoryear{{Garcia Lopez} \& {Spruit}}{{Garcia Lopez} \&
  {Spruit}}{1991}]{1991ApJ...377..268G}
{Garcia Lopez} R.~J.,  {Spruit} H.~C.,  1991, \mn@doi [\apj] {10.1086/170356},
  \href {http://ukads.nottingham.ac.uk/abs/1991ApJ...377..268G} {377, 268}

\bibitem[\protect\citeauthoryear{{Goldreich} \& {Kumar}}{{Goldreich} \&
  {Kumar}}{1990}]{1990ApJ...363..694G}
{Goldreich} P.,  {Kumar} P.,  1990, \mn@doi [\apj] {10.1086/169376}, \href
  {http://ukads.nottingham.ac.uk/abs/1990ApJ...363..694G} {363, 694}

\bibitem[\protect\citeauthoryear{{Goldreich}, {Murray}  \& {Kumar}}{{Goldreich}
  et~al.}{1994}]{1994ApJ...424..466G}
{Goldreich} P.,  {Murray} N.,   {Kumar} P.,  1994, \mn@doi [\apj]
  {10.1086/173904}, \href {http://adsabs.harvard.edu/abs/1994ApJ...424..466G}
  {424, 466}

\bibitem[\protect\citeauthoryear{{Hurlburt}, {Toomre}  \&
  {Massaguer}}{{Hurlburt} et~al.}{1986}]{1986ApJ...311..563H}
{Hurlburt} N.~E.,  {Toomre} J.,   {Massaguer} J.~M.,  1986, \mn@doi [\apj]
  {10.1086/164796}, \href {http://adsabs.harvard.edu/abs/1986ApJ...311..563H}
  {311, 563}

\bibitem[\protect\citeauthoryear{{Kumar}, {Talon}  \& {Zahn}}{{Kumar}
  et~al.}{1999}]{1999ApJ...520..859K}
{Kumar} P.,  {Talon} S.,   {Zahn} J.-P.,  1999, \mn@doi [\apj]
  {10.1086/307464}, \href {http://adsabs.harvard.edu/abs/1999ApJ...520..859K}
  {520, 859}

\bibitem[\protect\citeauthoryear{{Lecoanet} \& {Quataert}}{{Lecoanet} \&
  {Quataert}}{2013}]{2013MNRAS.430.2363L}
{Lecoanet} D.,  {Quataert} E.,  2013, \mn@doi [\mnras] {10.1093/mnras/stt055},
  \href {http://ukads.nottingham.ac.uk/abs/2013MNRAS.430.2363L} {430, 2363}

\bibitem[\protect\citeauthoryear{{Montalban}}{{Montalban}}{1994}]{1994A&A...281..421M}
{Montalban} J.,  1994, \aap, \href
  {http://ukads.nottingham.ac.uk/abs/1994A%26A...281..421M} {281, 421}

\bibitem[\protect\citeauthoryear{{Montalban} \& {Schatzman}}{{Montalban} \&
  {Schatzman}}{1996}]{1996A&A...305..513M}
{Montalban} J.,  {Schatzman} E.,  1996, \aap, \href
  {http://ukads.nottingham.ac.uk/abs/1996A%26A...305..513M} {305, 513}

\bibitem[\protect\citeauthoryear{{Montalb{\'a}n} \&
  {Schatzman}}{{Montalb{\'a}n} \& {Schatzman}}{2000a}]{2000A&A...354..943M}
{Montalb{\'a}n} J.,  {Schatzman} E.,  2000a, \aap, \href
  {http://ukads.nottingham.ac.uk/abs/2000A%26A...354..943M} {354, 943}

\bibitem[\protect\citeauthoryear{{Montalb{\'a}n} \&
  {Schatzman}}{{Montalb{\'a}n} \& {Schatzman}}{2000b}]{montalban2000a}
{Montalb{\'a}n} J.,  {Schatzman} E.,  2000b, \aap, \href
  {http://adsabs.harvard.edu/abs/2000A%26A...354..943M} {354, 943}

\bibitem[\protect\citeauthoryear{{Paxton}, {Bildsten}, {Dotter}, {Herwig},
  {Lesaffre}  \& {Timmes}}{{Paxton} et~al.}{2011}]{2011ApJS..192....3P}
{Paxton} B.,  {Bildsten} L.,  {Dotter} A.,  {Herwig} F.,  {Lesaffre} P.,
  {Timmes} F.,  2011, \mn@doi [\apjs] {10.1088/0067-0049/192/1/3}, \href
  {http://adsabs.harvard.edu/abs/2011ApJS..192....3P} {192, 3}

\bibitem[\protect\citeauthoryear{Phillips}{Phillips}{1966}]{phillips1966dynamics}
Phillips O.,  1966, The Dynamics of the Upper Ocean.
Cambridge monographs on mechanics and applied mathematics, Cambridge U.P., \url
  {https://books.google.co.uk/books?id=2H1RAAAAMAAJ}

\bibitem[\protect\citeauthoryear{{Pin{\c c}on}, {Belkacem}  \&
  {Goupil}}{{Pin{\c c}on} et~al.}{2016}]{pincon2016a}
{Pin{\c c}on} C.,  {Belkacem} K.,   {Goupil} M.~J.,  2016, \mn@doi [\aap]
  {10.1051/0004-6361/201527663}, \href
  {http://adsabs.harvard.edu/abs/2016A%26A...588A.122P} {588, A122}

\bibitem[\protect\citeauthoryear{{Press}}{{Press}}{1981}]{1981ApJ...245..286P}
{Press} W.~H.,  1981, \mn@doi [\apj] {10.1086/158809}, \href
  {http://ukads.nottingham.ac.uk/abs/1981ApJ...245..286P} {245, 286}

\bibitem[\protect\citeauthoryear{{Press} \& {Rybicki}}{{Press} \&
  {Rybicki}}{1981}]{1981ApJ...248..751P}
{Press} W.~H.,  {Rybicki} G.~B.,  1981, \mn@doi [\apj] {10.1086/159199}, \href
  {http://ukads.nottingham.ac.uk/abs/1981ApJ...248..751P} {248, 751}

\bibitem[\protect\citeauthoryear{{Ramiaramanantsoa} et~al.,}{{Ramiaramanantsoa}
  et~al.}{2018}]{2018MNRAS.480..972R}
{Ramiaramanantsoa} T.,  et~al., 2018, \mn@doi [\mnras] {10.1093/mnras/sty1897},
  \href {http://adsabs.harvard.edu/abs/2018MNRAS.480..972R} {480, 972}

\bibitem[\protect\citeauthoryear{{Rieutord} \& {Zahn}}{{Rieutord} \&
  {Zahn}}{1995}]{rieutord1995a}
{Rieutord} M.,  {Zahn} J.-P.,  1995, \aap, \href
  {http://adsabs.harvard.edu/abs/1995A%26A...296..127R} {296, 127}

\bibitem[\protect\citeauthoryear{{Rogers} \& {Glatzmaier}}{{Rogers} \&
  {Glatzmaier}}{2005}]{2005MNRAS.364.1135R}
{Rogers} T.~M.,  {Glatzmaier} G.~A.,  2005, \mn@doi [\mnras]
  {10.1111/j.1365-2966.2005.09659.x}, \href
  {http://adsabs.harvard.edu/abs/2005MNRAS.364.1135R} {364, 1135}

\bibitem[\protect\citeauthoryear{{Rogers} \& {McElwaine}}{{Rogers} \&
  {McElwaine}}{2017}]{2017arXiv170904920R}
{Rogers} T.~M.,  {McElwaine} J.~N.,  2017, preprint, \href
  {http://adsabs.harvard.edu/abs/2017arXiv170904920R} {} (\mn@eprint {arXiv}
  {1709.04920})

\bibitem[\protect\citeauthoryear{{Rogers}, {Lin}  \& {Lau}}{{Rogers}
  et~al.}{2012}]{2012ApJ...758L...6R}
{Rogers} T.~M.,  {Lin} D.~N.~C.,   {Lau} H.~H.~B.,  2012, \mn@doi [\apjl]
  {10.1088/2041-8205/758/1/L6}, \href
  {http://ukads.nottingham.ac.uk/abs/2012ApJ...758L...6R} {758, L6}

\bibitem[\protect\citeauthoryear{{Rogers}, {Lin}, {McElwaine}  \&
  {Lau}}{{Rogers} et~al.}{2013}]{2013ApJ...772...21R}
{Rogers} T.~M.,  {Lin} D.~N.~C.,  {McElwaine} J.~N.,   {Lau} H.~H.~B.,  2013,
  \mn@doi [\apj] {10.1088/0004-637X/772/1/21}, \href
  {http://adsabs.harvard.edu/abs/2013ApJ...772...21R} {772, 21}

\bibitem[\protect\citeauthoryear{{Schatzman}}{{Schatzman}}{1993}]{1993A&A...279..431S}
{Schatzman} E.,  1993, \aap, \href
  {http://ukads.nottingham.ac.uk/abs/1993A%26A...279..431S} {279, 431}

\bibitem[\protect\citeauthoryear{{Shiode}, {Quataert}, {Cantiello}  \&
  {Bildsten}}{{Shiode} et~al.}{2013}]{2013MNRAS.430.1736S}
{Shiode} J.~H.,  {Quataert} E.,  {Cantiello} M.,   {Bildsten} L.,  2013,
  \mn@doi [\mnras] {10.1093/mnras/sts719}, \href
  {http://adsabs.harvard.edu/abs/2013MNRAS.430.1736S} {430, 1736}

\bibitem[\protect\citeauthoryear{{Staquet} \& {Sommeria}}{{Staquet} \&
  {Sommeria}}{2002}]{2002AnRFM..34..559S}
{Staquet} C.,  {Sommeria} J.,  2002, \mn@doi [Annual Review of Fluid Mechanics]
  {10.1146/annurev.fluid.34.090601.130953}, \href
  {http://ukads.nottingham.ac.uk/abs/2002AnRFM..34..559S} {34, 559}

\bibitem[\protect\citeauthoryear{Sutherland}{Sutherland}{2010}]{sutherland2010internal}
Sutherland B.,  2010, Internal Gravity Waves.
Cambridge University Press, \url
  {https://books.google.co.uk/books?id=KNmaBAAAQBAJ}

\bibitem[\protect\citeauthoryear{{Tayar} \& {Pinsonneault}}{{Tayar} \&
  {Pinsonneault}}{2013}]{2013ApJ...775L...1T}
{Tayar} J.,  {Pinsonneault} M.~H.,  2013, \mn@doi [\apjl]
  {10.1088/2041-8205/775/1/L1}, \href
  {http://adsabs.harvard.edu/abs/2013ApJ...775L...1T} {775, L1}

\bibitem[\protect\citeauthoryear{{Townsend}}{{Townsend}}{1966}]{townsend1966a}
{Townsend} A.~A.,  1966, \mn@doi [Journal of Fluid Mechanics]
  {10.1017/S0022112066000661}, \href
  {http://adsabs.harvard.edu/abs/1966JFM....24..307T} {24, 307}

\bibitem[\protect\citeauthoryear{{Zahn}, {Talon}  \& {Matias}}{{Zahn}
  et~al.}{1997}]{1997A&A...322..320Z}
{Zahn} J.-P.,  {Talon} S.,   {Matias} J.,  1997, \aap, \href
  {http://adsabs.harvard.edu/abs/1997A%26A...322..320Z} {322, 320}

\makeatother
\end{thebibliography}


\bsp	
\label{lastpage}
\end{document}